\documentclass[aps,prd,twocolumn,superscriptaddress]{revtex4}
\usepackage{graphicx}
\usepackage{amsmath}
\usepackage{amsmath,amsfonts,amssymb}
\usepackage{pdfpages}
\usepackage{graphics}
\usepackage{epsfig}
\usepackage{amsmath,amsfonts,amssymb}
\usepackage{braket}
\usepackage{mathbbol}
\usepackage[T1]{fontenc}
\usepackage{bbold}
\usepackage{amsmath}
\usepackage{amsmath,amsfonts,amssymb}
\usepackage{times}
\usepackage{braket}
\usepackage[T1]{fontenc}
\usepackage{bbold}
\usepackage{mathbbol}
\usepackage{graphicx}
\usepackage{times}
\DeclareGraphicsExtensions{.png,.pdf}

\usepackage{color}

\begin{document}
\baselineskip=12pt
\def\black{\textcolor{black}}
\def\red{\textcolor{black}}
\def\blue{\textcolor{blue}}
\def\green{\textcolor{black}}
\def\be{\begin{equation}}
\def\ee{\end{equation}}
\def\bea{\begin{eqnarray}}
\def\eea{\end{eqnarray}}
\def\orc{\Omega_{r_c}}
\def\om{\Omega_{\text{m}}}
\def\E{{\rm e}}
\def\bearst{\begin{eqnarray*}}
\def\eearst{\end{eqnarray*}}
\def\peleven{\parbox{11cm}}
\def\peffec{\peight{\bearst\eearst}\hfill\peleven}
\def\pspace{\peight{\bearst\eearst}\hfill}
\def\ptwelve{\parbox{12cm}}
\def\peight{\parbox{8mm}}

\title{Matrix formalism of excursion set theory: \\ A new approach to statistics of dark matter halo counting}


\author{Farnik Nikakhtar}
\email{nikakhtar-AT-physics.sharif.edu}
\author{Shant Baghram}
\email{baghram-AT-sharif.edu}
\address{Department of Physics, Sharif University of Technology, P.~O.~Box 11155-9161, Tehran, Iran}


\begin{abstract}
Excursion set theory (EST) is an analytical framework to study the large-scale structure of the Universe. EST introduces a procedure to calculate the number density of structures by relating the cosmological linear perturbation theory to the nonlinear structures in late time.
In this work, we introduce a novel approach to reformulate the EST in matrix formalism. We propose that the matrix representation of EST will facilitate the calculations in this framework.
The method is to discretize the two-dimensional plane of variance and density contrast of EST, where the trajectories for each point in the Universe lived there. The probability of having a density contrast in a chosen variance is represented by a probability ket.  Naturally, the concept of the transition matrix pops up to define the trajectories. We also define the probability transition rate which is used to obtain the first up-crossing of trajectories and the number count of the structures. In this work we show that the discretization let us study the non-Markov processes by forcing them to look like a Wiener process. Also we discuss that the zero drift processes with Gaussian and also non-Gaussian initial conditions can be studied by this formalism. The continuous limit of the formalism is discussed, and the known Fokker-Planck dispersion equation is recovered. Finally we show that the probability of the most massive progenitors can be extracted in this framework.
\end{abstract}


\maketitle

\section{Introduction} \label{sec:intro}

The cosmological large scale structure (LSS) data will be the dominant source of information for future studies in cosmology. Cosmological deep questions can be addressed by LSS data, especially the questions which are related to dark universe [the physics of dark matter (DM) and dark energy (DE) \cite{Weinberg:2012es}].  The cosmic structures are formed via the growth of the primordial seeds of density contrast (gravitational potential perturbations) seen as temperature anisotropies in cosmic microwave background radiation (CMB) via gravitational instability \cite{Dodelson2003}. The cosmological perturbation theory is the key framework to address the linear regime structure formation in the Universe \cite{Ma:1995ey}. The gravitational instability brings the linear regime to an end and we find patches of the Universe that become more and more dense which must be described by nonlinear physics. Galaxies, groups of galaxies and galaxy clusters are manifestations of the nonlinear structures. Besides the nonlinear baryonic structures that we see in electromagnetic wavelength, the LSS has a dark matter counterpart as well. The angular power spectrum of CMB temperature anisotropies and matter density power spectrum indicate that the DM is the backbone of structures in the Universe. There are observational evidences that the DM  is distributed in halo shape structures which are the host of baryonic matter \cite{Bertone:2004pz}.
Now we should note that, the nonlinear gravitational physics of structure formation changes the arena and the questions. The interesting observable parameters  become the statistic of structures (DM halos) like the probability distribution function (PDF) of halos in terms of their mass. These statistics are linked to the mass (luminosity) profile of galaxies. The distribution of dark matter tracers (like galaxies and 21 cm emission \cite{Lidz:2013tra}) has a major role in establishing the standard model of cosmology known as the $\Lambda$CDM model \cite{Tegmark:2003ud}.

In order to find the distribution of nonlinear structures, a long path is passed. Spherical collapse \cite{Gunn:1972sv} and Press-Schechter (PS) formalism  \cite{Press:1973iz} start the venue, where later this approach is developed in a stochastic framework known as excursion set theory (EST) (or extended PS formalism) \cite{Bond:1990iw}. The EST is based on a probabilistic approach to the structure formation. The evolution of density contrast with respect to the density contrast variance in different smoothing regions is studied. In this context, the trajectory of density contrast versus variance for different points of the Universe is obtained. The trajectories that pass a specified critical density contrast become qualified to be nonlinear structures (halos of DM) which must become the host of baryonic matter. In other words, DM halos are related to the regions where the trajectories crossed the barrier. There is an idea of reformulating the EST with the path integral approach by \cite{Maggiore:2009rv} which helps to investigate the extensions of the theory \cite{Maggiore:2009rw,Maggiore:2009rx,Adshead:2012hs}.

In this work, we propose a new framework to study the EST approach of structure formation know as matrix formalism of excursion set theory (MF-EST).  The idea that we develop in this work becomes from discretization of the trajectories, where we will be capable of defining probability kets and transition matrix for any desired trajectory.
We obtained all the plausible results of standard EST for Markov processes. However we show that this can be done by the concepts of transition matrix and the probability transition rate. This will open up  a new horizon to study the deviations from the standard case of EST via mathematics of matrices. The discretization lets us study the stochastic processes of barrier crossing in the Markov limit. Also we are capable to study non-Markov cases via the transition matrix. Also as a specific example we study a non-Gaussian example and we show how MF-EST can be used to obtain the statistics of the first up-crossing in this nonstandard case. This formalism is suitable for data analysis studies, which tends to study the N-body simulations. The finite number of probability states in matrix formalism EST is an advantage for this type of studies.

The structure of the work is as follows: In Sec. \ref{sec:Overview} we review the nonlinear structure formation in the context of PS formalism and the standard excursion set theory as the extension of PS. In Sec. \ref{sec:Matrix}, we introduce the matrix formalism of excursion set theory and we discuss the theoretical basis of the framework. In Sec. \ref{sec:number}, we will discuss the first up-crossing concept and we find the number density of structures in this framework for a Wiener process. In this section we also compare the results with dark-sky simulations, where we used the top-hat filter, which makes the process non-Markov. In Sec. \ref{sec:ng}, we study the first up-crossing problem for a non-Gaussian initial condition as a specific example which deviates from the standard EST. In Sec. (\ref{sec:cont}) we discuss the continuous limit of matrix formalism and we will discuss the probability transition rate matrix and the derivation of the Fokker-Planck equation. In Sec. \ref{sec:mmp} we express the physics of the main progenitors in halo merger history in matrix formalism and finally in Sec. \ref{sec:conclusion} we conclude  and indicate the future prospects of this work. In addition, the Appendix of this paper shows the statistics in matrix formalism of EST and the capability of this formalism in calculating the statistical quantities.


\section{Theoretical Overview: Press-Schechter Formalism and Excursion Set Theory}

\label{sec:Overview}
One of the main topics of interest in nonlinear structure formation is the study of the number density of structures in the Universe.
In the linear regime we have the density contrast field as $\delta ({\bf{x}},t)$ defined for each point in the Universe. In this regime the density of structures grows linearly with growth function as $\delta({\bf{x}},t)=D(t)\delta_0({\bf{x}})$, where $\delta_0$ is the linearly extrapolated present time density contrast and $D(t)$ is the linear growth function normalized to unity in present time  \cite{Bernardeau:2001qr,Cooray:2002dia}. Now the idea is that by using the spherical collapse, (top-hat) distribution of the matter \cite{Gunn:1972sv} which introduces a critical density $\delta_c\simeq 1.68$ \footnote{We should note that $\delta_c$ is the extrapolation of the density contrast of a collapsed structure in linear regime.}, we are able to find the number of bound structures. The regions with $\delta({\bf{x}},t)>\delta_c$ collapsed at time $t$ and after virialization they become a dark matter halo which will be capable to host a baryonic structure \cite{Cooray:2002dia}. Note that the barrier crossing condition can be reexpressed as $\delta_0 > \delta_c(t)$, where $\delta_c(t) = \delta_c / D(t)$, which means the density contrast is extrapolated to the present time and the barrier is moved corresponding to the time that we are interested in studying the nonlinear structure. In this context in order to think about the collapsed regions Press and Schechter \cite{Press:1973iz} considered a smoothed field as below
\begin{equation}
\delta_s({\bf{x}};R)\equiv \int \delta_0({\bf{x^\prime}})W({\bf{x}}+{\bf{x}}';R)d^3{\bf{x}}',
\end{equation}
where subindices refer to smoothed density function, $W({\bf{x}}+{\bf{x}}';R)$ is the window function with a specific radius of smoothing $R$. The shape of the window function is arbitrary, we can use the top-hat filter in real space for example as it is introduced in PS formalism. For each collapsed region, we can assign a mass $M=\gamma_f\bar{\rho}R^3$, where $\bar{\rho}$ is the background density and $\gamma_f$ is the volume coefficient which depends on the choice of the window function (i.e. $\gamma_f=4\pi/3$ for top-hat window function). The important ansatz of the PS formalism  is that the probability of the regions which satisfy the condition of $\delta_s > \delta_c(t)$ is the same as the fraction of the volume at time $t$, contain halos with mass greater than $M$. We should note that this specific mass is related to the smoothing scale $R$ due to choice of the window function. \\
One important point to indicate is that we can assume (eligibly), $\delta_0$ is a Gaussian function, accordingly $\delta_s$ is Gaussian as well.
In this case the probability of finding regions with mass larger than $M$ will be:
\begin{equation}
{\cal{P}}(>\delta_c(t))=\frac{1}{\sqrt{2\pi}\sigma (M)} \int_{\delta_c(t)}^{\infty}e^{-{\delta^2_s}/{2\sigma^2(M)}}d\delta_s.
\end{equation}
The distribution of density contrast in the equation above is related to cosmological linear theory by the variance of fluctuations defined as below:
\begin{equation}
S(M) \equiv \sigma ^2(M) = \langle \delta_s^2({\bf{x}};R) \rangle = \frac{1}{2\pi^2} \int_0^{\infty} P(k)\tilde{W}^2(kR)k^2dk,
\end{equation}
where $\tilde{W}$ is the Fourier transform of the window function and $P(k)$ is the present-time $z=0$ power spectrum of matter perturbations in linear regime, in which the physics of dark matter and dark energy plays an essential role in its construction. A very important point to indicate is that if we use the matter power spectrum of CDM we find that the variance is a monotonic decreasing function of mass $M$ (or equivalently the corresponding radius $R$).
The PS formalism has its own shortcomings and limitations. The important one is the cloud-in-cloud problem. The miscounting of the structures in PS formalism results in a miscalculated factor 2.
An alternative approach to the clustering of density contrast is the excursion set theory  defined by Bond {\it{et al.}} \cite{Bond:1990iw}. This theory  with a reasonable way solves the cloud-in-cloud problem. In this picture for each point of the Universe ${\bf{x}}$, there is a function of smoothed density contrast in terms of variance $\delta_s =\delta_s(S)$ \cite{Zentner:2006vw}. As we mentioned the monotonic behavior of $S =S(M)$ let us set the $S$  equivalent to the smoothing radius $R$ and the mass enclosed in it. Note that in the limit of $S \rightarrow 0$ which is equal to $M\rightarrow \infty$ the smoothing box is equal to the whole Universe and the density contrast is equal to zero. Increasing $S$ corresponds to decreasing the weighting radius (or mass), in this case we anticipate that the $\delta_s(S)$ deviates from zero.
One crucial point to add to this picture is usage of the sharp k-space filter defined as below:
\begin{equation}
\delta_s({\bf{x}};R)=\int d^3k\tilde{W}(kR)\delta_{k,0}e^{i{\bf{k}}.{\bf{x}}}=\int_{k<k_c}d^3k\delta_{k,0}e^{i{\bf{k}}.{\bf{x}}},
\end{equation}
where $k_c = 1/R$ is the size of top-hat filter in k-space and $\delta_{k,0}$ is the Fourier modes of density contrast. The advantage of this specific filter is that the $\Delta\delta_s$ which is the change in the density contrast when we move from $k_c$ to $k_c+\Delta k_c$ is a Gaussian random field. In addition this process becomes a Markov one. It means that these processes have a memory of only its previous step \cite{Maggiore:2009rv}. The variance of Gaussian random variable is $\langle (\Delta \delta_s)^2 \rangle = \sigma^2 (k_c+\Delta k_c) - \sigma ^2 (k_c)$, where $\sigma^2(k_c)$ is defined as
\be
\sigma^2(k_c)=\frac{1}{2\pi^2}\int_{k<k_c}P(k)k^2.
\ee
Accordingly the value of $\delta_s$ at a given position in the Universe is performing a Markovian random walk. We should note that, if we change the filter function the process will not be a Markovian.
The Markovian behavior of trajectories with the sharp k-space filter in the EST plane is crucial, as  the number density of the structures can be obtained analytically by solving the cloud-in-cloud problem. We should note that the sharp-k space filter is not a realistic window function for mass estimation in simulations and observations. In the case, if we use more realistic window functions, such as real space top-hat or a Gaussian filter functions, the excursion process of $\delta$ versus variance deviates from Markovianity.". To address this problem, some extensions of EST such as correlated steps walks are proposed \cite{Musso:2011ck,Musso:2012qk,Musso:2013pha,Musso:2013pja,Musso:2014jda}. However in this work, we developed the new formalism in the context of the Markovianity, where we can force the non-Markov process to look like a Markov one by discretizing the processes with Markov length. This will be one of the nice features which appeared in matrix formalism.

In the EST, we find the first up-crossing (FU) using the idea of mirroring trajectories as
\be \label{eq:upcross}
F_{FU}(>S_*) = \int _{-\infty}^{\delta_c}\left[{\cal{P}}(S_*,\delta_s) - {\cal{P}}( S_*, 2\delta_c - \delta_s)\right]d\delta_s,
\ee
where $F_{FU}(>S_*)$ shows the probability of the trajectories which will have their first up-crossing in larger variance than the specific $S_*$. The second term in Eq.(\ref{eq:upcross}) comes from the mirroring trajectories. For every trajectory that passes the barrier in smaller variances ($S<S_*$) but in $S_*$ are below the barrier in point $(S,\delta_s)$ there is a mirroring trajectory that at the specific point it has the value of $(S, 2\delta_c - \delta_S)$. This procedure can be applied in EST formalism, whenever the corresponding walks are Markov. Accordingly, the first up-crossing can be rewritten as below:
\be \label{eq:upcross1}
F_{FU}(>S_*) = \int _{-\infty}^{\delta_c}{\cal{P}}(S_*,\delta_s)d\delta_s - \int_{\delta_c}^{\infty} {\cal{P}}(S_*,\delta_s)d\delta_s.
\ee
There is a notation that is used to show the first up-crossing probability in variance interval of $(S, S + dS)$ which is defined as $f_{FU} = - \partial F / \partial S$. The idea discussed above will be used in the next sections for the matrix formalism of EST.
The number density of structures in the mass range of $M$ and $M+dM$ can be obtained from the first up-crossing distribution as
\be \label{nFU}
n(M,t)dM=\frac{\bar{\rho}}{M}\frac{\partial F_{FU}(>M)}{\partial M}dM= \frac{\bar\rho}{M}\frac{\partial F_{FU}}{\partial S}|\frac{dS}{dM}|dM,
\ee
where $\bar{\rho}$ is the background matter density, and $n(M,t)$ is the number density of the structures in the mass range of $M$ and $M+dM$. In case we use the Gaussian probability function the number density becomes
\begin{equation}
n(M,t)dM= \sqrt{\frac{2}{\pi}}\frac{\bar{\rho}}{M^2}\frac{\delta_c}{\sigma(M)}\exp(-\frac{\delta_c^2}{2\sigma^2(M)})|\frac{d\ln \sigma}{d\ln M}|dM.
\end{equation}
The main advantage of using the EST formalism is its ability to construct the merger history of the dark matter halos \cite{Kauffmann:1993gv,Sheth:1998ay,Somerville:1997df,Cole:2000ex,Bosch:2001ej}. For this task, the essential quantity to calculate is the main progenitors of dark matter halos, which can be obtained via conditional probability in EST. Assume that we have a halo of mass $M_2$ at time $t_2$; the probability that this halo's main progenitor has a mass $M_1$ and it is merged in time $t_1$ can be calculated with conditional first up-crossing probability as
\begin{eqnarray}
f_{FU}(S_1, \delta _1 | S_2 , \delta _2 ) dS_1 &=& \frac{1}{\sqrt{2\pi}} \frac{\delta_1 - \delta _2}{ (S_1 - S_2 ) ^{3/2}} \\ \nonumber
&\times&\exp [- \frac{(\delta_1 - \delta _2)^2}{2(S_1 - S_2)}] dS_1
\end{eqnarray}
where $S_1$ and $S_2$ are variances related to mass $M_1$ and $M_2$ and $\delta_1$ and $\delta_2$ .

A final piece to add to this theoretical overview section is the introduction of the diffusion-type equation. The probability ${\cal{P}}(S,\delta_s)$ of a trajectory in a specific variance, to be in $\delta_s$ and $\delta_s + \Delta\delta_s$, is obtained from the Fokker-Planck equation below\cite{Mo2010}:
\be
\frac{\partial {\cal{P}}(S,\delta_s)}{\partial S} = - \frac{\partial (\mu{\cal{P}}(S,\delta_s))}{\partial \delta_s} + \frac{1}{2}\frac{\partial^2(\Sigma^2 {\cal{P}}(S,\delta_s))}{\partial \delta_s^2},
\ee
where $\mu\equiv \lim _{\Delta S \to 0} \frac{\langle \Delta \delta_s | \delta _s \rangle }{\Delta S}$ and $\Sigma^2 \equiv \lim _{\Delta S\rightarrow 0}\frac{\langle (\Delta \delta_s)^2 | \delta _s \rangle }{\Delta S}$ are drift and diffusion parameters respectively. For the sharp k-space filter, we have $\mu=0$ and $\Sigma^2=1$. Accordingly the Fokker-Planck  equation reduces to  \cite{Maggiore:2009rv}
\be
\frac{\partial{\cal{P}}(S,\delta_s)}{\partial S} = \frac{1}{2}\frac{ \partial ^2 {\cal{P}}(S,\delta_s) }{\partial \delta_s^2}.
\ee
A very important point worth stating it again is that in the case of Markovianity (sharp k-space filter), when $\Delta S \rightarrow 0$, the variance of changes in density contrast become equal to steps in variance axis $\langle (\Delta \delta _s )^2 \rangle / \Delta S \rightarrow 1$.

\begin{table}
  \centering
\begin{tabular}{|c|c|c|c|}
  \hline
    \hline
Type of processes & Initial condition &  Drift  & MF-EST \\
\hline
Markov & Gaussian &  Zero drift & \checkmark \\
Markov & Non-Gaussian & Zero drift & \checkmark\\
Markov & Non- Gaussian & With drift & $\times$ \\
Non-Markov & Gaussian & Zero drift & $\rightarrow$ Markov \\
Non-Markov & Non-Gaussian & Zero drift & $\rightarrow$ Markov \\
Non-Markov & Non-Gaussian & With drift & $\times$ \\
\hline
\end{tabular}
\caption{The EST processes are categorized due to status of their Markovianity-initial condition and drift. In the last row we show the scope of MF-EST.} \label{table1}
\end{table}

In Table \ref{table1}, we categorize the EST processes, with their essential characteristics. In general the processes can be divided to Markov and non-Markov cases which depends on the filter function which we use to smooth the density contrast. The Matrix formalism of EST which we represent in this work can change the non-Markov processes to Markov ones, by setting the discretization steps equal to the Markov length. The Markov length is defined  as a lag in a time series, in which the process looks like Markov. However, we should note that we lose information due to coarse graining.
The other characteristics of the EST processes are the initial condition, which can be divided to Gaussian and non-Gaussian cases. Finally the drift of the processes can be an important characteristic.
In this work, we assert that by the concept of discretization, we can always force the process to look like a Markovian one. The number density of structures can also be obtained when the drift is zero, with both Gaussian and non-Gaussian initial conditions. The nonzero drift solutions of the Fokker-Planck  equation is not the subject of this work. In the next section we will introduce our new framework of excursion set theory. \\ \\

\section{The Matrix Formalism of EST}
\label{sec:Matrix}
In this section we first present the matrix formulation of excursion set theory. In the first subsection we come up with the representation of formalism. In the second subsection we discuss the processes of construction of the transition matrix for a random walk with equal probability. In the third subsection we discuss the construction of transition matrix with a Gaussian distribution and finally we wrap up this section by construction of the transition matrix by using the EST trajectories for a Wiener process.

\subsection{ EST representation in matrix representation}

As we discussed in the first section, the excursion set theory in its essence has a stochastic point of view to the theory of structure formation.
For each point of the Universe we can compute the density contrast in a box of size $R$. The size of the box (window function) has a one-to-one relation with variance $S$.  By changing the size of the window function, we have new variance and density contrast, accordingly we have a trajectory in a two-dimensional space of density contrast $\delta_m$ [the subscript $m$ indicates that we are dealing with the total matter (CDM + baryonic) density contrast and it is the same smoothed density contrast shown as $\delta_s$ in the previous section.] and variance $S$, for each position in the Universe. Hereafter $(S,\delta_m)$ will be called the {\it{2D-plane of EST}}.
Considering all different locations in the Universe, we will have many trajectories in the $(S,\delta_m)$ plane.
The physical quantities of our interest are the probabilities. This means that we are interested to know what fraction of the trajectories in a specific time slice (redshift) passes from a specific $(S_*,\delta^*_{m})$ in 2D-plane of EST.
With this idea in mind, in order to present the matrix formalism, we assume that the density contrast can be discretized by a large\footnote{Here by large we mean a thermodynamical large number where we can construct the Gaussian profiles and transitions. It is possible to choose a maximum and minimum density contrast due to the simulation (observation) which is under study and then we can discretized the probability ket in sufficient large number.} finite number of  probability states for each $S$. Now it makes sense to define the probability of each density contrast in terms of a vector in any desired variance as below:
\begin{equation}
	\Ket{\mathcal{P}_s} =\left(\begin{smallmatrix}
		\mathcal{P}_s (\delta_{m}^1) \\
		\mathcal{P}_s (\delta_{m}^2) \\
		\vdots \\
		\mathcal{P}_s (\delta_{m}^n)
	\end{smallmatrix} \right)\\,
\end{equation}
where subscript $s$ in the {\it{probability ket}} $\Ket{\mathcal{P}_s}$ indicates that the vector is defined for a specific size of a window function corresponding to variance $S$.
Each component of the vector ${\cal{P}}_s(\delta_m^i)$ shows the normalized probability $(S ,\delta_m^i)$ in all trajectories.
This interpretation of the probability vector can be written in terms of basis vectors as below:
\begin{equation} \label{eq:ket}
	\Ket{\mathcal{P}_s} = \displaystyle\sum_{i} \mathcal{P}_s (\delta_{m}^i) \ket{\delta_{m}^i},
\end{equation}\\
where the basis vectors $\ket{\delta_{m}^i}$ are defined as below:
\begin{equation*}
	\ket{\delta_{m}^1} =
	\left(\begin{smallmatrix}
		1 \\
		0 \\
		\vdots \\
		0
	\end{smallmatrix}\right),
	\hspace{15 pt}
	\ket{\delta_{m}^2} =
	\left(\begin{smallmatrix}
		0 \\
		1 \\
		\vdots \\
		0
	\end{smallmatrix}\right),
	\hspace{15 pt} \dots \hspace{25 pt}
	\Ket{\delta_{m}^n} =
	\left(\begin{smallmatrix}
		0 \\
		0 \\
		\vdots \\
		1
	\end{smallmatrix}\right).
\end{equation*}
Now it is obvious that the orthonormality condition becomes
\begin{equation}
	\Braket{\delta_m^i | \delta_m^j} = \delta_{ij},  \hspace{35 pt} \displaystyle\sum_{i} \Ket{\delta_m^i}\Bra{\delta_m^i} = {{1}}.
\end{equation}
Accordingly it is easy to think that by knowing the probability ket, $\ket{{\cal{P}}_s}$, we can obtain the probability of finding the trajectory in point $(S,\delta^k_m)$ ($k$ indicates a specific desired density contrast). Due to the fact that the sum of probabilities for a trajectory is equal to unity in a specific variance we have
\begin{equation}
	\displaystyle\sum_{i} \Braket{\delta_m^i | \mathcal{P}_s} = 1.\\
\end{equation}
Following the spirit of the EST, now we want to study the physics of the trajectories.
Accordingly the next step will be the implementation of Markovianity as a first simplified assumption in matrix formalism of excursion set theory. This means that the value of a point trajectory in the $S-\delta_m$ plane depends only to the last step of its excursion. This can be formalized through the  ${\cal{T}}_{ij}$ coefficients as below
\begin{equation} \label{eq:transfer1}
	\mathcal{P}_s (\delta_m^i) = \displaystyle\sum_{j} \mathcal{T}_{ij} \mathcal{P}_{s-\Delta s} (\delta_m^j), \\
\end{equation}\\
where the sum is considered to be applied over all possible transitions from step $S-\Delta S$ to $S$, where $\Delta S$ is the step in the S-axis; after this we can set it to unity for convenience in notation.
We should keep in mind that $\Delta S$ in this formalism is strictly related to the resolution of the size of the window function, which  we can apply in N-body simulations or in LSS surveys. We will discuss it further in the conclusion and the future prospect section.

\begin{figure}[t!]
\center
\includegraphics[width=0.5\textwidth ]{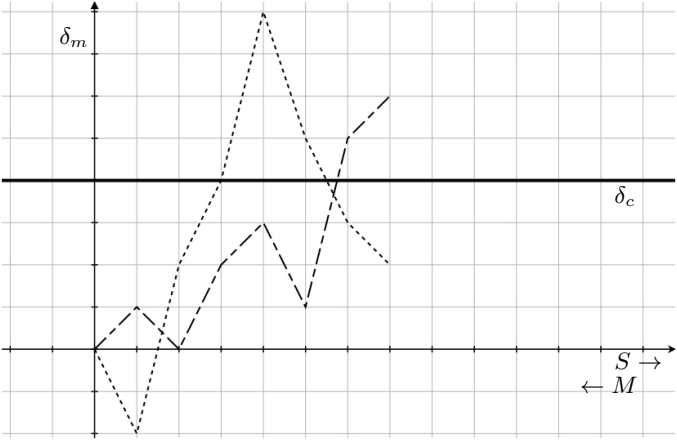}
\put(-233,165){$\delta_m$}
\put(-23,105){$\delta_c$}
\put(-23,42){$S\rightarrow$}
\put(-36,33){$\leftarrow M$}
\caption{The trajectories in the two-dimensional plane of EST $(S,\delta_m)$ are plotted, the x-axis is the variance and y-axis is the density contrast. The plane is discretized and trajectories our obliged to be on discrete points. The solid line indicates the critical density contrast $\delta_c$.}
\label{fig-dis}
\end{figure}

In Fig. \ref{fig-dis}, we plot a schematic figure of the 2D plane of EST. The plane is discretized in density contrast and variance, and each trajectory is related to a point in the Universe. The positions of the trajectory are obtained by setting $\delta_m$ as a density contrast for a smoothing window function with a size of  $R$ correspondingly, the variance(the x-axis of the plane). The horizontal solid line shows a critical density that the first up-crossing of the trajectory indicates the size of the structure. \\
The discrete nature of the 2D plane of EST and the probability states lead us to the idea of the {\it{transition matrix}}. The transition matrix  ${\cal{T}}_{ij}$ is defined naturally as below:
\begin{equation}  \label{eq:transfer2}
\mathcal{T}_{ij} = \Braket{\delta_m^i | \hat{\mathcal{T}} | \delta_m^j}	= \mathcal{P}( S ,\delta_m^i  |  S-1, \delta_m^j ),
\end{equation} \\
where the second equality is the conditional probability which relates the points of the $(S-1,\delta_m^j)$ and $(S,\delta_m^i)$. (Note that $\Delta S = 1$ is used interchangeably.)
Accordingly using Eqs.(\ref{eq:transfer1}) and (\ref{eq:transfer2}) we can find the probability of finding the trajectory in $(S,\delta_m^i)$ as below
\begin{equation}
	\Braket{\delta_m^i | \mathcal{P}_s} = \displaystyle\sum_{j} \Braket{\delta_m^i | \hat{\mathcal{T}} | \delta_m^j} \Braket{\delta_m^j | \mathcal{P}_{s-1}}
	= \Braket{\delta_m^i | \hat{\mathcal{T}} | \mathcal{P}_{s-1}}.
\end{equation}\\
This means that for knowing the probability ket in each point of the EST plane we should know the transition matrix and the probability in the previous step:\\
\begin{equation}\label{eq:Ts}
	\Ket{\mathcal{P}_s} = \hat{\mathcal{T}} \Ket{\mathcal{P}_{s-1}}.
\end{equation}
Now by assuming that the Markovianity is working all the way from the biggest box  with the variance of $S_0\equiv 0$ until the desired point in the 2D plane of EST, and the fact that the transition matrix is independent of variance ( this assumption is equal to the assumption of the homogeneity in time, which means that we are dealing with a stationary stochastic process), we can construct the probability ket in step $S$ just by knowing the transition matrix and initial probability ket $\Ket{\mathcal{P}_0}$ as below:
\begin{equation}
	\boxed{\Ket{\mathcal{P}_s} = \hat{\mathcal{T}}^{s} \Ket{\mathcal{P}_0}}
\end{equation}
We should emphasis here that we set the steps in variance to unity as it mentioned but in general to reach to a specific step of $S$ we need $n$ steps where $S = n\times \Delta S$. This is clear and easy but important to note because in upcoming discussions we will use the $S$ and $n$ interchangeably.
The independency of the transition matrix from variance is a simplified assumption for structure formation in the language of EST.
In general the transition matrix can depend on the value of the variance, accordingly we can obtain the probability ket in the specific variance $S$ with the relation below:
\begin{equation}
	\Ket{\mathcal{P}_s} = \mathcal{T}(S) \mathcal{T}(S-\Delta S) \dots \mathcal{T}(S-n\Delta S) \Ket{\mathcal{P}_0}.
\end{equation}
It should be mentioned that we have another assumption in this work, which is homogeneity in density contrast. It means the transition of trajectories just depends on the density contrast changes $\Delta\delta_m$ rather than the value of $\delta_m$ in steps of transition. So in the context of matrices, we should use {\it{Toeplitz}} matrices or diagonal-constant matrices, in which each descending or ascending diagonal, from down to up has a constant value.
The cornerstone of the matrix formalism of EST is the construction of the transition matrix, which encapsulates in it the physics of excursion. Any deviation from the simplified assumption of structure formation, like  sharp k-space filter and Markovianity, can be reflected in the construction of this matrix, keeping in mind that working with matrices can simplify the problem in the computational sense.\\
In the next subsections we discuss the construction of the transition matrix with equal probability, Gaussian probability transitions, and from exact trajectories respectively.

\subsection{Construction of transition matrix for a random walk with equal probability}
In this section we want to construct the transition matrix for a simple example of random walk with an equal probability transition.
This will be done by  using the analogy of the creation and annihilation operators in quantum mechanics.
In this framework we define $(\hat{a}_{A})_{ij}$ as the ascending transition matrix similar to creation operator which augments the $\delta_m$ one state as the variance is changed by one step. In other words, we need an operator to set
\be
{\cal{P}}_s(\delta^i_m) \equiv {\cal{P}}_{s-1}(\delta_m^{i+1}).
\ee
Note that superindices are assigned to elements of matrices and vectors in a standard way, that is why the ascending transition matrix decreases the index of a specific element. This can be done by transition matrix $\hat{a}_A$ as
\be
{\cal{P}}_s(\delta_m^i)= \sum_j (\hat{a}_A)_{ij}{\cal{P}}_{s-1} (\delta_m^j).
\ee
Now it is straightforward that the ascending parameter will become Kronecker delta,
\be
(\hat{a}_A)_{ij}=\delta^k_{i+1,j},
\ee
where superscript $k$ indicates the Kronecker delta.
The matrix representation of $(\hat{a}_{A})_{ij}$  will be

\begin{equation*}
	\hat{a}_A =
	 \left(\begin{smallmatrix}
   0 & 1 & 0   & \hdots    & 0 \\
   \vdots & \ddots   & \ddots  & \ddots & 0 \\
   \vdots &    & \ddots & \ddots & 1  \\
   \vdots &    &        & \ddots & 0 \\
   0 & 0  & \hdots  & 0 &  0
  \end{smallmatrix}\right).
\end{equation*}\\
It is straightforward to define the  descending transition matrix $(\hat{a}_{D})_{ij}$, which should set ${\cal{P}}_s(\delta^i_m) \equiv {\cal{P}}_{s-1}(\delta_m^{i-1})$ as below:
\begin{eqnarray}
(\hat{a}_D)_{ij}=\delta^k_{i-1,j},
\end{eqnarray}
where the matrix representation is
\begin{equation*}
	\hat{a}_D =
	 \left(\begin{smallmatrix}
   0 & \hdots & \hdots & \hdots & 0 \\
   1 & \ddots   &   &  & \vdots \\
   0 & \ddots   & \ddots  &   & 0  \\
   \vdots & \ddots   & \ddots   & \ddots & 0 \\
   0 & \hdots & 0 & 1 &  0
  \end{smallmatrix}\right).
\end{equation*}

Now we can write a desired transition matrix as a linear combination of $\hat{a}_A$ and $\hat{a}_D$. For the simplest case consider that with half amplitude the trajectory goes upward or downward equally as below
\begin{equation}
	\hat{\mathcal{T}} = \frac{1}{2}(\hat{a}_A + \hat{a}_D).
\end{equation}
Now with the assumption of Markovianity for the trajectories, the $\hat{\mathcal{T}}^s$  become a matrix which is constructed by $s$ times of transition. It is applied to the initial probability ket in the exact number of times that we need to arrive at point $S$ as below:
\begin{equation}
	\hat{\mathcal{T}}^s = \frac{1}{2^s} \displaystyle\sum_{k=0}^s \binom sk (\hat{a}_A)^k (\hat{a}_D)^{s-k}.
\end{equation}
Now by applying the above transition matrix to the initial probability state we will find the final probability ket whose elements are the coefficients of the binomial distribution.
In another words, each specific  probability state in the 2D plane of EST  is accessible  by the ascending and descending matrix (operators).
The matrix formalism of EST presents the finite transition matrices which can be diagonalized and represented in terms of its eigenvectors and eigenvalues as
\begin{equation}
	\hat{\mathcal{T}} = \displaystyle\sum_{k} \lambda_k \Ket{\lambda_k}\Bra{\lambda_k}.
\end{equation}
Now it is crystal clear that the $s$th power of a diagonal matrix is
\begin{equation}
	\hat{\mathcal{T}}^{s} = \displaystyle\sum_{k} \lambda_k^s \Ket{\lambda_k}\Bra{\lambda_k},
\end{equation}
where the eigenvectors satisfy the orthonormality condition as below
\begin{equation}
	\Braket{\lambda_i|\lambda_j} = \delta_{i,j}^k , \hspace{25pt} \displaystyle\sum_i \ket{\lambda_i} \bra{\lambda_i} = 1.
\end{equation}
The above construction shows why the matrix formalism of EST is useful to make all the physics of the random walk expressed in the transition matrix.
As a final word to this subsection, we present two theorems which help to complement the working mechanism of the formalism. \\

{\it{Theorem one}}. If  $\hat{{\cal{T}}}$ is an $n\times n$ stochastic matrix (in which the rows sum to unity) then $\lambda=1$ is an eigenvalue.\\
To prove the first theorem, we define the {\it{adder Delta}} $\Bra{\Delta}$ \footnote{The terminology of adder Delta is borrowed from electronic and computer science. Adder is a digital circuit that performs addition of numbers.
Adder Delta in this context has the same function.} as a useful tool for summing up the probability states.
According to the definition of the $\hat{\mathcal{T}}$ matrix and the conditional probability we have
\begin{equation}\label{eq:Delta}
	\displaystyle\sum_{i} \Braket{\delta_m^i | \hat{\mathcal{T}} | \delta_m^j} = \displaystyle\sum_{i} \mathcal{P}(S, \delta_m^i  | S-1, \delta_m^j ) = 1.
\end{equation}
Now we can define the adder Delta $\Bra{\Delta_m^j}$ as
\begin{equation}
	\displaystyle \sum_{j} \Bra{\delta_m^j} = \Bra{\Delta}.
\end{equation}
A very interesting point is that the  adder Delta $\Bra{\Delta}$ is the eigenvector of the transition matrix with eigenvalue of unity
\begin{equation} \label{eq:Delta1}
	\Bra{\Delta}\hat{\mathcal{T}} = \displaystyle\sum_{i,j} \Bra{\delta_m^i}\hat{\mathcal{T}} \Ket{\delta_m^j}\Bra{\delta_m^j}= \displaystyle\sum_{j} \Bra{\delta_m^j} = \Bra{\Delta}.
\end{equation}

{\it{Theorem two}}. If $\lambda$ is a complex eigenvalue of a stochastic matrix $\hat{{\cal{T}}}$, then $|\lambda| \leq 1$.\\
The second theorem  indicates  the presence of the maximum eigenvalue for the transition matrix and the fact that all the other eigenvalues are smaller than unity.
Since $\Bra{\lambda_i}\hat{\mathcal{T}} = \lambda_i \Bra{\lambda_i}$, for each value of $\delta_m^\alpha$ we have
\begin{equation}\label{eq:lambdai}
	\braket{\lambda_i|\hat{\mathcal{T}}|\delta_m^\alpha} = \lambda_i \Braket{\lambda_i|\delta_m^\alpha}, \hspace{10pt} \forall \alpha\\
\end{equation}
where we define $\Braket{\lambda_i|\delta_m^\alpha} = \xi_\alpha$. Accordingly adding a unity in Eq.(\ref{eq:lambdai}) we will have
\begin{equation}
	\displaystyle\sum_{j} \Braket{\lambda_i|\delta_m^j}\braket{\delta_m^j|\hat{\mathcal{T}}|\delta_m^\alpha} = \lambda_i \xi_\alpha .
\end{equation}
Now the absolute value of $|\lambda_i \xi_\alpha|$ can be written as
\begin{equation}
	|\lambda_i||\xi_\alpha| \leq \displaystyle\sum_{j} |\xi_j| |\braket{\delta_m^j|\hat{\mathcal{T}}|\delta_m^\alpha}|
	 \leq |\xi_{max}| \displaystyle\sum_{j} |\braket{\delta_m^j|\hat{\mathcal{T}}|\delta_m^\alpha}|,
\end{equation}
where the second inequality comes from replacing the  $|\xi_j|$ with its maximum value. Then we will have
\begin{equation}
	|\lambda_i| |\xi_\alpha| \leq |\xi_{max}|,
\end{equation}
Now if we set $\xi_\alpha = \xi_{max}$ we can show that all the eigenvalues are smaller that the unity
\begin{equation}
	 |\lambda_i| \leq 1 .
\end{equation}
This means that by increasing the steps in the variance axis, the effect of the eigenvalues that are smaller from unity become less and less, and the system approaches asymptotically to a stationary probability ket, which is identified with an eigenvector corresponding to an eigenvalue of unity.
In the next subsection, we study the transition matrix for a Gaussian case which is a step closer to the matrix representation of EST.

\subsection{Construction of transition matrix for a Gaussian case}


\begin{figure}[t!]
\center
\includegraphics[width=0.5\textwidth]{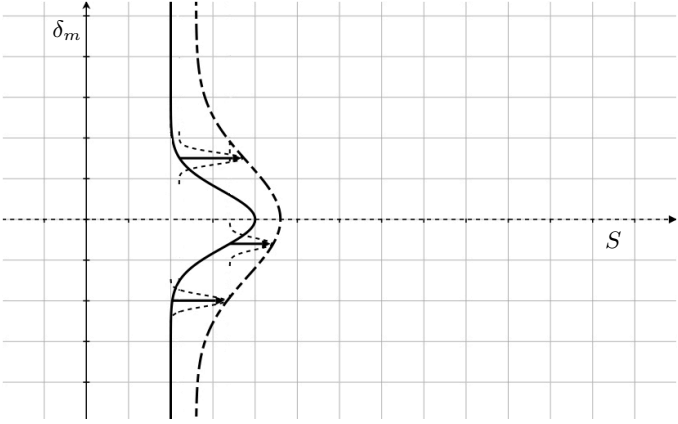}
\put(-236,164){$\delta_m$}
\put(-27,84){$S$}
\caption{The transition matrix reconstruction from adding Gaussian transition functions in $\Delta  \delta_m$. The figure shows that the steepest descent method adding the peak of transition functions will make the Gaussian PDF of density contrast for a specified variance. (The x-axis is the variance and the y-axis is the density contrast.) }
\label{fig:tra}
\end{figure}


One of the cornerstones of EST is the Gaussian transition matrix. This transition probability must be consistent with the Gaussian profile of density contrast for each specific variance. In this subsection we study this type of transition.
The transition matrix due to its definition can be written as below:
\begin{equation} \label{eq:gaussian}
	\hat{\mathcal{T}} = \displaystyle\sum_{u,d} \mathcal{T}_{ud}~ \hat{a}_A^u ~\hat{a}_D^d, \hspace{30pt} \mathcal{T}_{ud}=\mathcal{A}~ e^{-\frac{|u-d|^2}{2\sigma_\Delta^2}},
\end{equation}
where $\mathcal{A}$ is the normalization coefficient and $\sigma^2_{\Delta}$ is the variance of the Gaussian transition matrix [this variance is the analogue of $\langle (\Delta\delta)^2 \rangle$ introduced in Sec. \ref{sec:Overview}, as the mean of Gaussian transition is equal to zero]. It is worth mentioning that the transition matrix depends only to $\sigma^2_{\Delta}$ which is equal to the $\Delta S$ in sharp k-space filter in the limit of $\Delta S \rightarrow 0$. The  density contrast difference is shown symbolically as $\Delta\delta_m = u - d$. A very crucial property of the transition matrix introduced in Eq.(\ref{eq:gaussian}) is the symmetry of the coefficients of the ascending and descending transition matrices. This is the analogue to the Langevin equation with a white noise for a random walk motion  with zero drift in a 2D-EST plane \cite{Mo2010}, which is an essential piece for defining the mirror trajectories in EST.
It is an interesting question to ask how the transition matrix profile will look like after $n$ steps in the 2D-EST plane. To answer this question, we should keep in mind that $(\hat{a}_A)_{ij} = \delta_{(i+1),j}$ and $(\hat{a}_D)_{(i-1) ,j} = \delta_{(i+1),j}$, accordingly we can write the ascending and descending transition matrices by the basis vectors $\Ket{\delta_m^i}$ as
\begin{equation}
	\hat{a}_A = \displaystyle\sum_i \Ket{\delta_m^{i}}\Bra{\delta_m^{i+1}}, \hspace{35pt} \hat{a}_D = \displaystyle\sum_i \Ket{\delta_m^{i+1}}\Bra{\delta_m^{i}}.
\end{equation}
Accordingly the change due $\Delta\delta_m $ states, for the ascending and descending matrices, will lead us to write the transition matrix as
\begin{equation}
	\hat{\mathcal{T}} = \displaystyle\sum_{u,d,i} \mathcal{T}_{ud} \Ket{\delta_m^{i+u}} \Bra{\delta_m^{i+d}}.
\end{equation}
The transition matrix with $n$ steps is easy to construct now. As the transition matrix is variance independent, we are capable to obtain the Gaussian $\Ket{\mathcal{P}_s}$ from initial probability ket $\Ket{\mathcal{P}_0}$, where step $S$ is equal to $n$ times $\Delta S$ ($S=n\Delta S$)
\begin{equation}
	\hat{\mathcal{T}}^n = \mathcal{A}^n \displaystyle\sum_{\{u_i, d_i\}, i} e^{-\frac{\sum_i (\Delta\delta_m^i)^2}{2\sigma_\Delta^2}} \Ket{\delta_m^{i + u_n +\sum_{i=1}^{n-1} (\Delta\delta_m^i)^2}} \Bra{\delta_m^{i+d_n}}.
\end{equation}
Now the matrix product of $\hat{\mathcal{T}}^n$ on the initial probability ket $\Ket{\mathcal{P}_0}$ becomes
\begin{equation} \label{eq:pssteep}
	\Ket{\mathcal{P}_s} = \hat{\mathcal{T}}^n \Ket{\mathcal{P}_0}
 = \mathcal{A}^n \displaystyle\sum_{\{\Delta\delta_m^i \}}e^{-\frac{\sum_{i}(\Delta\delta_m^i)^2}{2\sigma_{\Delta}^2}} \Ket{\delta_m^{\sum_{i}\Delta\delta_m^i}}.
\end{equation}
\begin{figure}[!t]
\center
\includegraphics[width=0.5\textwidth]{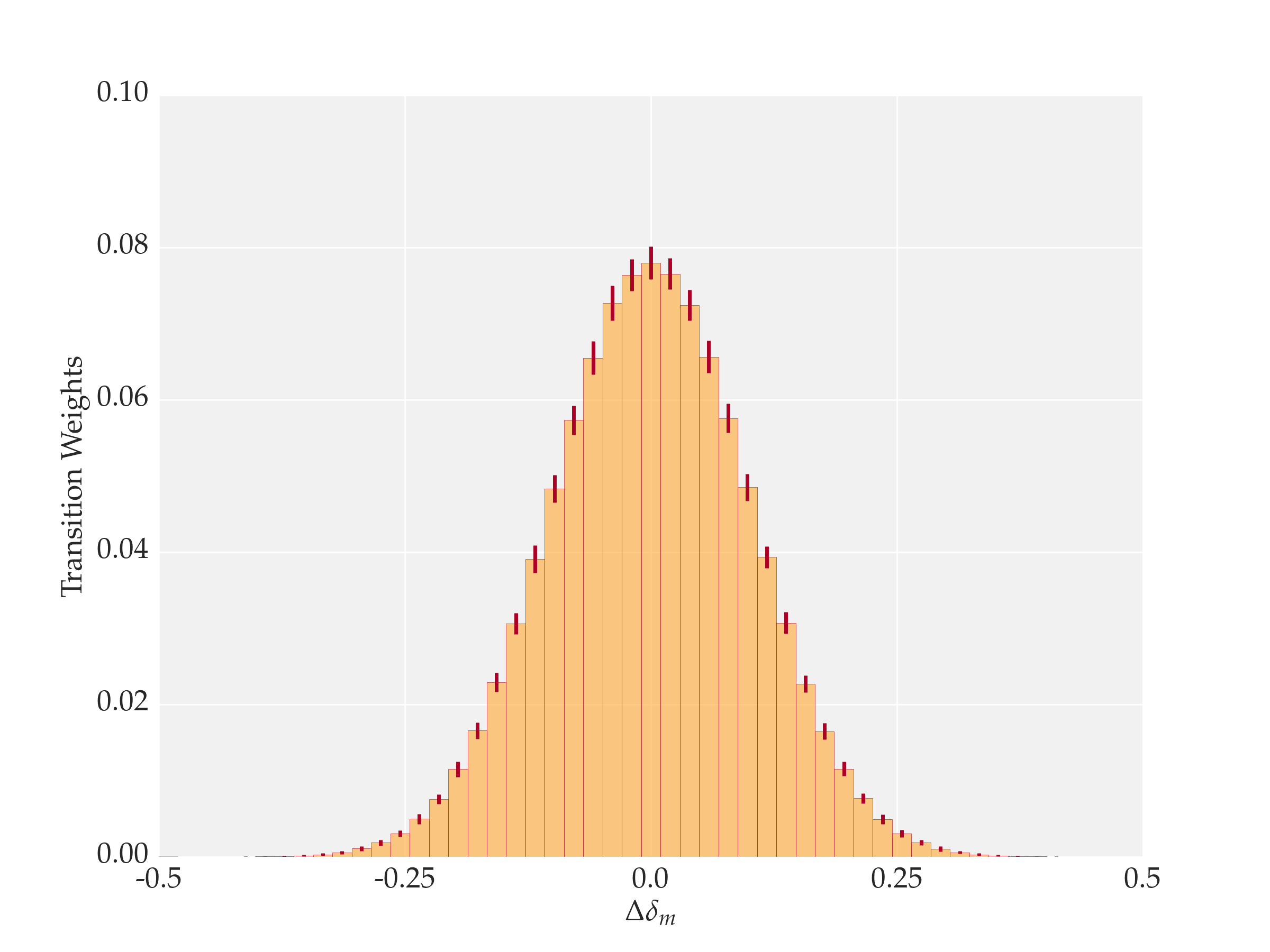}
\caption{The normalized histogram of the transition in $\Delta\delta_m$ for synthesized Wiener  trajectories. The error bars were obtained from 1$\sigma$ confidence level of an ensemble of 100 series with $10^3$ trajectories.}
\label{fig:WienerTra}
\end{figure}


Now the very crucial point here is that, we have $n$ density contrast difference $\Delta\delta_m^i$ with $(i=1,2,..,n)$ and we want to reach to a specified density contrast $\delta_m^*$ at the variance $S$, ( $\Delta\delta_m^1 + \Delta\delta_m^2 + \hdots + \Delta\delta_m^n = \delta_m^*$).
To address this inquiry we will use the steepest descent method. \footnote{An extension of Laplace method to approximate a summation by maximizing the contribution of each term. }\\
Accordingly in order to maximize the contribution of each term in the summation we have to minimize the numerator of the exponential term  by using the method of Lagrange multipliers,
\begin{eqnarray}
&&\frac{\partial}{\partial\Delta\delta_m^i}  \left [(\Delta\delta_m^1)^2 + (\Delta\delta_m^2)^2 + \hdots + (\Delta\delta_m^n)^2 \right]
 \\ \nonumber &-& \lambda (\Delta\delta_m^1 + \Delta\delta_m^2 + \hdots + \Delta\delta_m^n)=0,
\end{eqnarray}
so we will have $2 \Delta\delta_m^i - \lambda = 0$ and then $\Delta\delta_m^i = \lambda/2$. As we know if we want to minimize the sum of a series of  squared terms, with the condition that the sum of the terms  is a constant, we should put them equal to each other. This means that the $\Delta \delta_m$ relates to the step in variance axis
\begin{equation}
\Delta\delta_m^i = \frac{\delta_m^*}{n},
\end{equation}
where $n$ is the number of steps that we need to reach to a specific variance $S = n\Delta S$.
This procedure is schematically shown in Fig. \ref{fig:tra}.
Now the final task is to put the result from the steepest descent method obtained in Eq.(\ref{eq:pssteep}) to derive the very familiar result of PS formalism in the new format,
\begin{equation}
	\displaystyle\sum_i (\Delta\delta_m^i)^2 = \Delta S \frac{(\delta_m^*)^2}{S}.
\end{equation}
Again as we discussed before in the sharp k-space filter in the limit of $\Delta S \rightarrow 0$ we have $\sigma^2_{\Delta} / \Delta S \rightarrow 1 $, accordingly  $\Ket{\mathcal{P}_s}$ can be written in terms of basis vectors
\begin{equation}
\boxed{\Ket{\mathcal{P}_s} = \tilde{\mathcal{A}} \displaystyle\sum_{\delta_m^*} e^{-\frac{(\delta_m^*)^2}{2S }} \Ket{\delta^*_m}}
\end{equation}
where $\tilde{\mathcal{A}}$ is the normalization constant.
A final word that the summation of Gaussian transitions gives rise to a Gaussian PDF for density contrast with the variance of S as expected.  Fig. \ref{fig:tra} shows how the PDF of density contrast is changed with the variance. For small scales we have larger variance and accordingly we will have a broadened PDF of density contrast.


This means that the very familiar result of PS formalism is obtained from the matrix formalism of EST. The finite discretized nature of PDF of density contrast will be a potentially great opportunity to check the EST in numerical simulations to check the validity of the Gaussian transition matrix. In the next subsection we will discuss the construction of the transition matrix from trajectories.

\subsection{Construction of transition matrix from trajectories}
A profound way, that we suggest to make the transition matrices from data or synthesized trajectories works as below.
By the knowledge of the ensemble of trajectories in the EST 2D plane, in each discretized step (variance $S$), we can plot the histogram of the transitions in $\Delta\delta_m$ by going to the next step $S+\delta S$. The trajectories can be obtained from N-body simulations by choosing an appropriate window function. It is also possible to construct the trajectories from a Langevin equation. In EST as we mentioned, the density contrast variance obeys
 \be \label{eq:Wiener}
 \frac{\partial \delta}{ \partial S}=\eta (S),
 \ee
 where $\eta$ has a correlation $C(S,S')=\langle \eta(S)\eta(S') \rangle$. We know that in the standard Markov case $C(S,S')=\delta^D(S-S')$, which is known as the Wiener process as well. So in this case we can generate the trajectories, with a white noise.
 The transition matrix $T_{ij}$ is a $n\times n$ Toeplitz matrix, which means that the value of each diagonal is constant. An important point to indicate here is that the Toeplitz matrices are homogeneous in density contract changes. In other words the probability of having a specific transition $\Delta\delta_m$ is independent of given density contrast. The dimension of the transition matrix is fixed by the number of grids in a specific variance step. To find the distribution of diagonal values of the transition matrix, we should calculate density contrast changes for each trajectory in a variance-step interval. This distribution is equivalent to distribution of $\Delta\delta_m$, so we should construct the histogram of density contrast changes. This histogram can be used straightforwardly to construct the transition matrix by substituting the values of the $\Delta\delta_m$ histogram into each diagonal of the transition matrix. It should be mentioned that this method of constructing the transition matrices is completely independent of the distribution of probability kets.

\begin{figure}[!t]
\center
\includegraphics[width=0.5\textwidth]{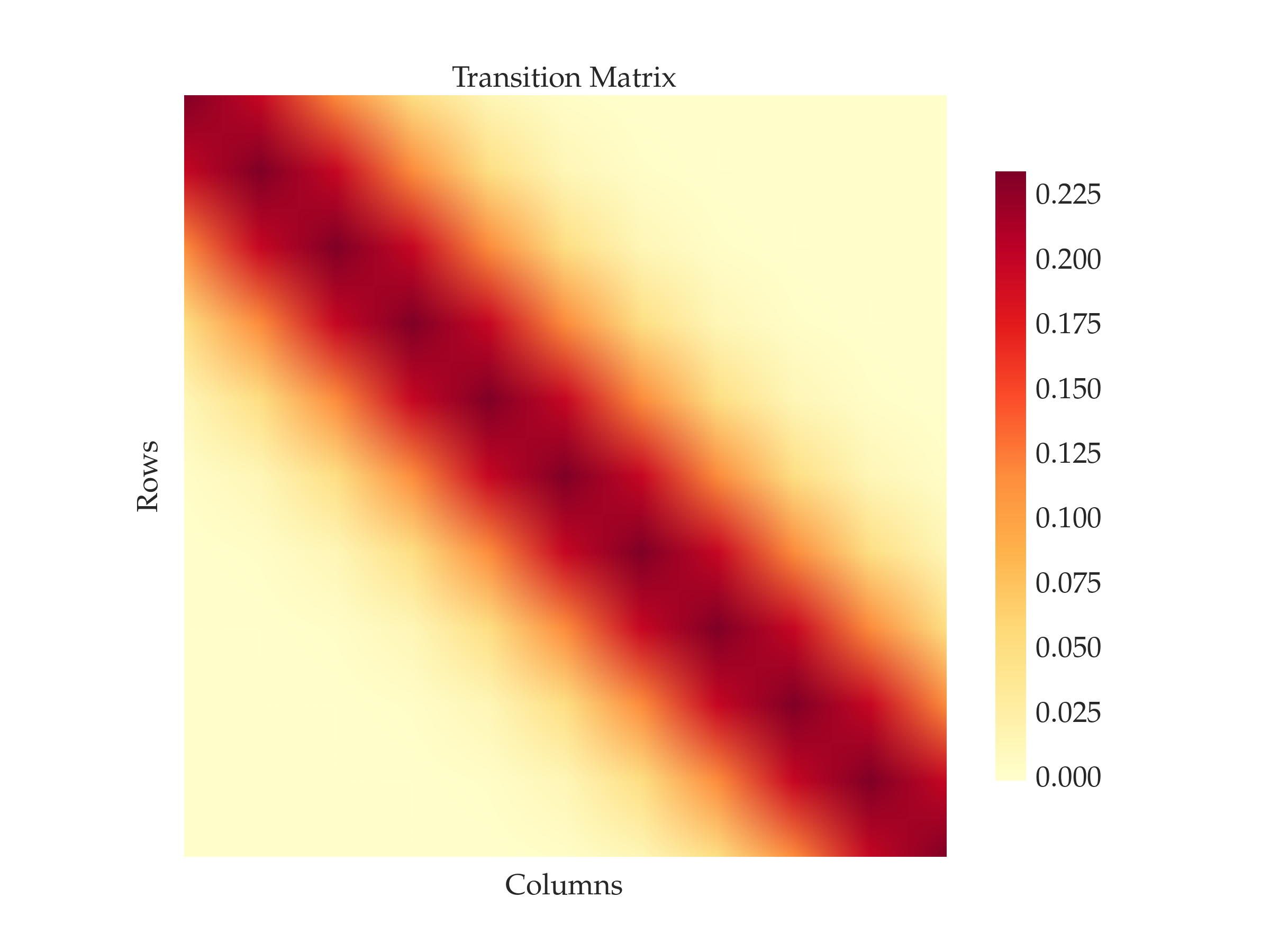}
\caption{The Toeplitz matrix of transition constructed for a Wiener process. The x and y axis represent the rows and columns of a transition matrix.}
\label{fig:toeplitz}
\end{figure}


In Fig. \ref{fig:WienerTra}, the normalized histogram of the transition in $\Delta\delta_m$ for synthesized Wiener  trajectories due to Eq.(\ref{eq:Wiener}) is plotted. The error bars were obtained from the 1$\sigma$ confidence level from an ensemble of 100 series of $10^3$ trajectories. In this plot, we assume the homogeneity of time. In Fig. \ref{fig:toeplitz}, we plot the Toeplitz matrix due to definition of the transition matrix. It is obvious that the main transition probability resides on the main diagonal of the transition. In the next section, we introduce the procedure which leads to the definition of first up-crossing.


\section {First up-crossing and  number count in the Matrix Formalism of EST }
\label{sec:number}
One of the main goals of EST is to find the mass (luminosity) profile of the nonlinear objects \cite{Schechter:1976iz}. This is done by an assumption that the number density of the structures with a specific mass (size of a window function or equivalently the mass variance) is proportional to the trajectories that pass the barrier of the critical density for the first time in the corresponding variance step. In the matrix formalism of EST this can be done by using the idea of counting the fraction of trajectories over a barrier. However, the crucial point arises when we want to address the traditional cloud-in-cloud problem shown in the PS formalism.
As discussed in Sec. \ref{sec:Overview}, in order to exclude (count) the substructures that reside in larger structures twice, we use the mirroring trajectory idea.
In order to apply the idea of counting the fraction of trajectories that pass the barrier with considering the mirror trajectories,  we define a block diagonal matrix ${\cal{\hat{M}}}$ respectively. For this task, we recall Eq.(\ref{eq:upcross1}), where we showed that the fraction of the first up-crossing trajectories is equal to the sum of the states from minus infinity to the critical density $\delta_c$ {\it{minus}} the sum of the states from critical density to infinity. Accordingly the matrix ${\cal{\hat{M}}}$ is
\begin{equation}
 \hat{\mathcal{M}}(\delta_c(z)) = \left(
    \begin{array}{r@{}c|c@{}l}
  &    \begin{smallmatrix}
        -1 & & 0 \\
          &\ddots&\\
        0 & & -1\rule[-1ex]{0pt}{2ex}
      \end{smallmatrix} & \mbox{\large0} & \\\hline
  &   \mbox{\large0}    & \begin{smallmatrix}
        1 & & 0 \\
          &\ddots&\\
        0 & & 1\rule[-1ex]{0pt}{2ex}
      \end{smallmatrix} &
    \end{array}
\right),
\end{equation}
where the cross point of the horizontal and vertical lines (the borders of the different partitions of matrix) is fixed by the position of the critical state $\delta_c$, which depends on the redshift.
It seems straightforward that by applying the block diagonal matrix $\hat{\mathcal{M}}$ on the probability ket, and using the adder Delta, we
can find the fraction of the trajectories that has their first up-crossing in matrix formalism
\be
F_{FU}(>S)=\Braket{\Delta| \hat{\mathcal{M}}|\mathcal{P}_s},
\ee
where $F(>S)$ is the fraction of trajectories that have their first up-crossing in $s>S$.
Obviously the next step must be the settlement of the relation between the fraction of first up-crossed trajectories with number density of structures. For this task the first up-crossing distribution $f_{FU}$ is obtained as
\begin{equation}
\boxed{	f_{FU}(S) = - \frac{\partial}{\partial S} \Braket{\Delta |\hat{\mathcal{M}} |\mathcal{P}_s}}
\label{Eq:f-fu}
\end{equation}
where the redshift dependence comes from the block diagonal matrix ${\cal{\hat{M}}}$. As the adder Delta $\Delta$ and matrix ${\cal{\hat{M}}}$ are both independent of variance, accordingly the number density of structures in the mass range of $M$ and $M+dM$ is as below:
\be
{n(M,z)dM = - \frac{\bar{\rho}}{M} \Braket{\Delta| \hat{\mathcal{M}} \frac{\partial}{\partial S} | \mathcal{P}_s} |\frac{dS}{dM}|dM}
\ee
This shows that in the case of the Gaussian PDF we will find the universality function for the structures \cite{Mo2010}.
The final piece of the number count is the introduction of the {\it{probability transition rate}} matrix. For this task we should calculate the variance derivative of probability ket
\begin{equation}
	\frac{\partial \Ket{\mathcal{P}_s}}{\partial S} = \frac{\Delta \ket{\mathcal{P}_s}}{\Delta s} = \frac{\Ket{\mathcal{P}_{s+\Delta s}} - \Ket{\mathcal{P}_s}}{\Delta S},
\end{equation}
where $\Delta S$ is the small step in variance axis (which can be set to unity). On the other hand, we can use the transition matrix to relate the two states $\Ket{\mathcal{P}_{s+\Delta s}} = \hat{\mathcal{T}} \Ket{\mathcal{P}_s},$. In this case the derivative of the probability ket becomes
\begin{equation}
	\frac{\Delta \ket{\mathcal{P}_s}}{\Delta S} = \frac{(\hat{\mathcal{T}} -\hat{{\bf{1}}})}{\Delta S} \Ket{\mathcal{P}_s}.
\end{equation}
The equation above shows that ${\Delta \ket{\mathcal{P}_s}}/{\Delta s}$ will be obtained by the {\it{probability transition rate}} $\hat{\mathcal{R}}$ and $\Ket{\mathcal{P}_s}$ as a simple relation
\be
\frac{\Delta \ket{\mathcal{P}_s}}{\Delta S} = \hat{\mathcal{R}} \Ket{\mathcal{P}_s},
\ee \label{eq:rate}
where probability transition rate is defined as
\begin{equation}
	\boxed{\hat{\mathcal{R}} = \frac{(\hat{\mathcal{T}} -\hat{{\bf{1}}})}{\Delta S}}
\label{Eq:rate}
\end{equation}
It means that by knowing the transition matrix ${\cal{\hat{T}}}$ we can construct the probability transition rate matrix.
We should note that the solution suggested for the first up-crossing by the procedure above works for a standard EST which is identified by the Gaussian Markov random process with zero drift. However our formalism can be extended to the non-Gaussian cases which preserve the symmetry of the trajectories around the average. This symmetry will allow us to use the concept of the block diagonal ${\cal{M}}$ matrix and adder Delta to obtain the first up-crossing distribution. In what follows we present standard examples of EST in matrix formalism. We devote a complete separate section on the development of theory in the non-Gaussian case. In the two upcoming subsections we investigate the first up-crossing distribution in for a Wiener process and also in dark sky simulation.


\subsection{First up-crossing in a Wiener process}
In this subsection, we use the Markov process to show that how the MF-EST works. For this task we generate the trajectories using the Langevin equation. Now we can construct the probability ket by assuming an initial probability  which is in a single state of $(S,\delta)=(0, 0)$. Then we apply the Gaussian transition matrix on this initial state to obtain the probability ket in the sequence of variance steps. The transition matrix is constructed by the histogram of $\Delta \delta_m$ which is represented in Fig. \ref{fig:WienerTra} and discussed in the previous section.
\begin{figure}[h!]
\minipage{0.4\textwidth}
  \includegraphics[width=\linewidth]{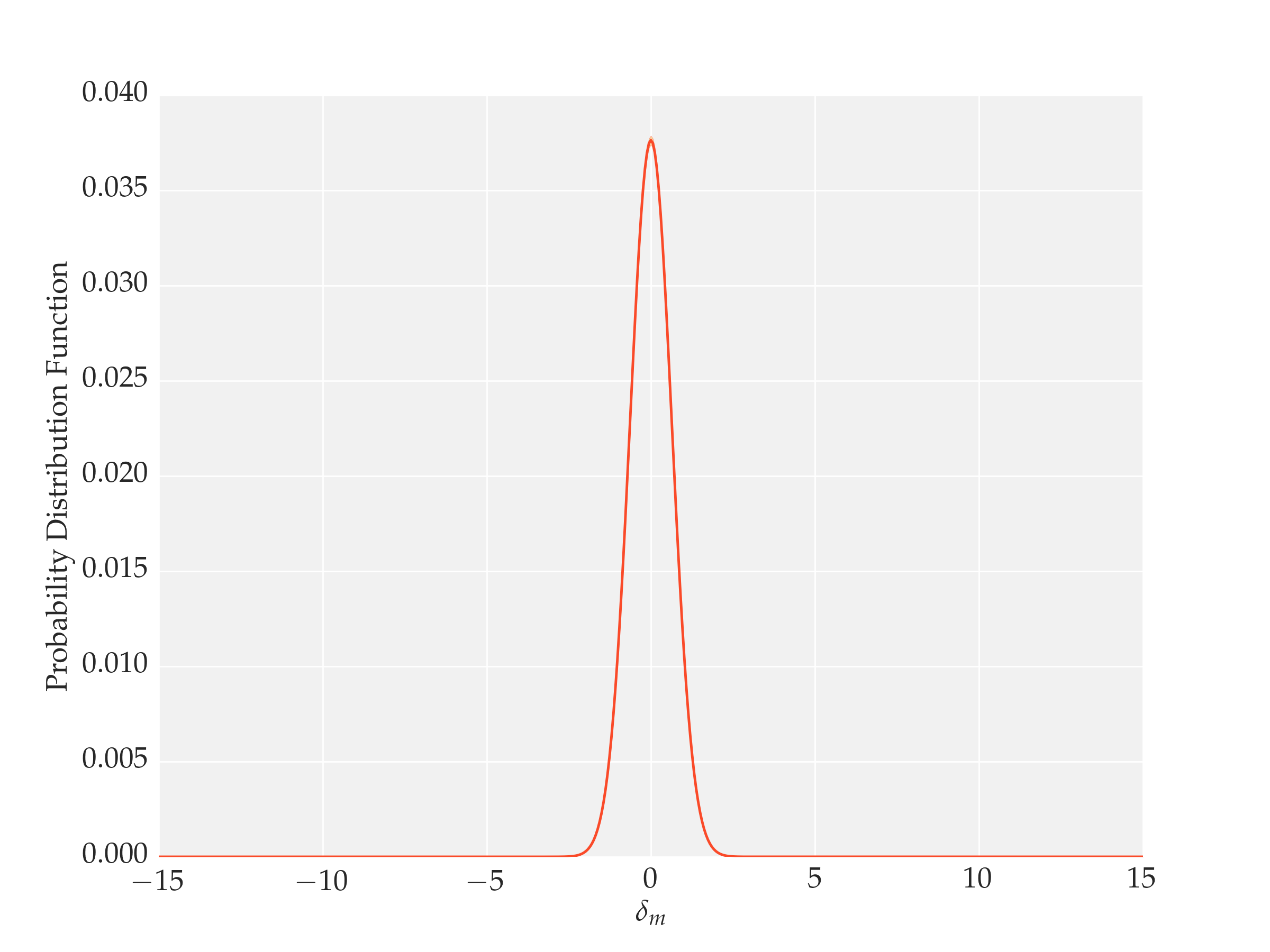}
\endminipage\hfill
\minipage{0.4\textwidth}
  \includegraphics[width=\linewidth]{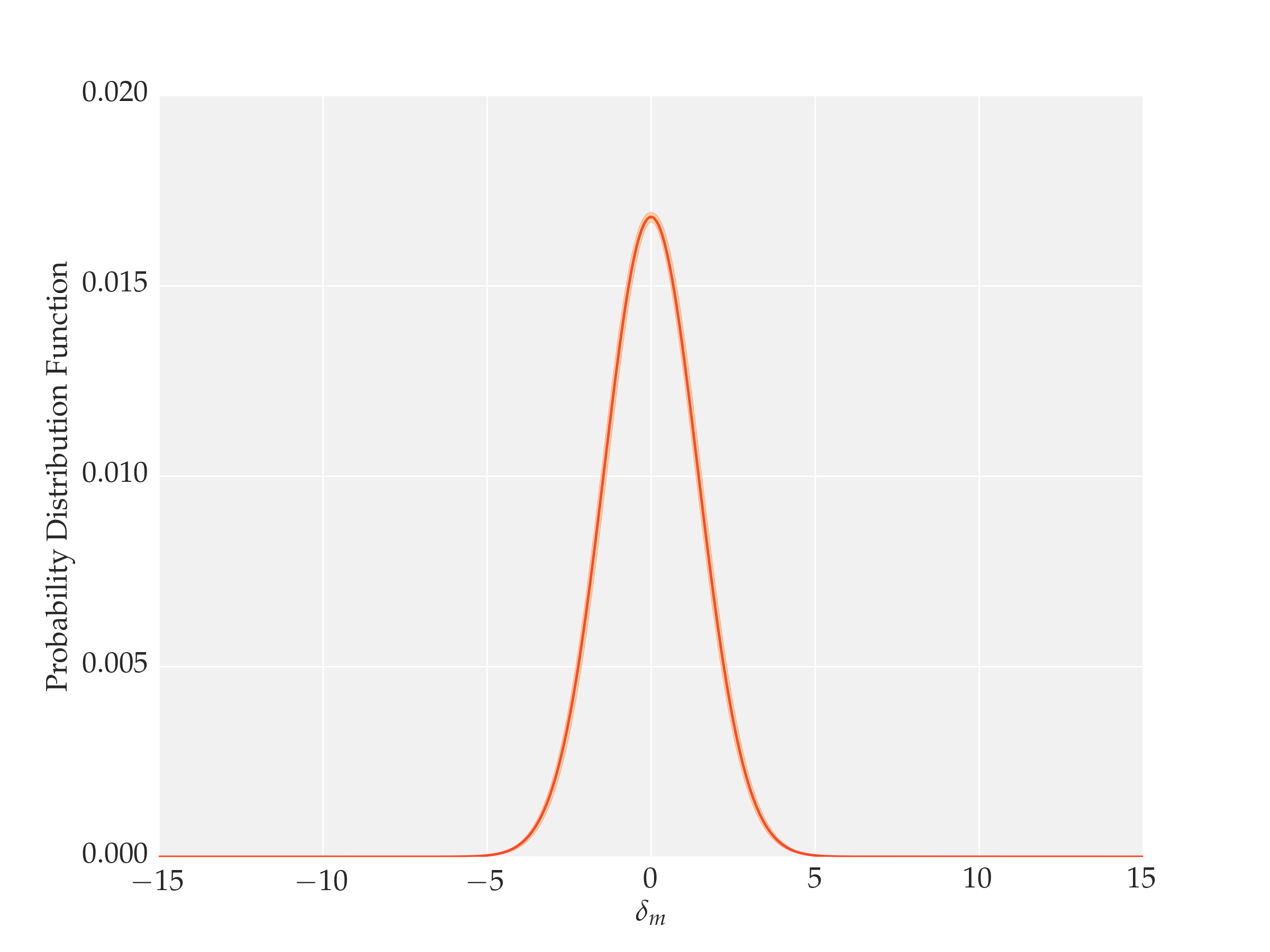}
\endminipage\hfill
\minipage{0.4\textwidth}%
  \includegraphics[width=\linewidth]{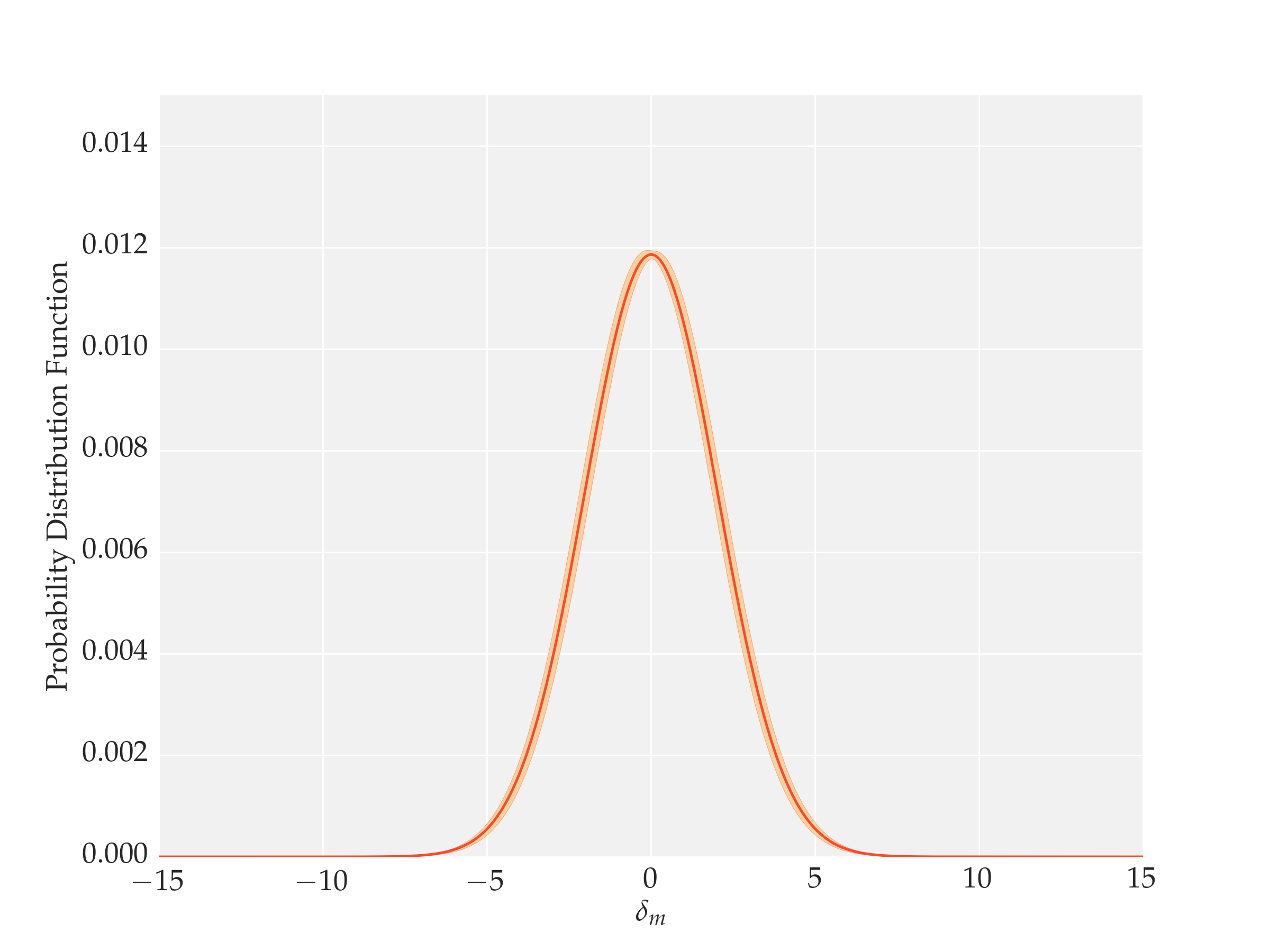}
\endminipage
\caption{From up to down, these figures show the normalized probability distribution function (PDF) of density contrast in three sequences of variances of $S=0.1$, $S=0.5$ and $S=1.0$ from a Wiener process. The error bars are 1$\sigma$ confidence level of the probability obtained from an ensemble of transition matrices introduced in Fig. \ref{fig:WienerTra}.} \label{fig:WPS}
\end{figure}
In Fig.(\ref{fig:WPS}), from up to down, we plot the normalized probability distribution function (PDF) of density contrast in three sequences of variances of $S=0.1$, $S=0.5$ and $S=1.0$ from a Wiener process. The error bars are 1$\sigma$ confidence level of the probability obtained from an ensemble of transition matrices introduced in Fig. \ref{fig:WienerTra}.
As we discussed, by increasing the steps in the x-axis of the 2D plane of EST, the variance of density contrast PDF is increased as well. Now by using the probability ket in any desired variance and the definition of transition rate matrix in Eq.(\ref{Eq:rate}), we can obtain the probability of first crossing. Note that hereafter we set $\Delta = S  0.01$ and discretization is done by resolution of $S/\Delta S$, which we keep the same hereafter.
\begin{figure}[h]
\center
\includegraphics[width=\linewidth]{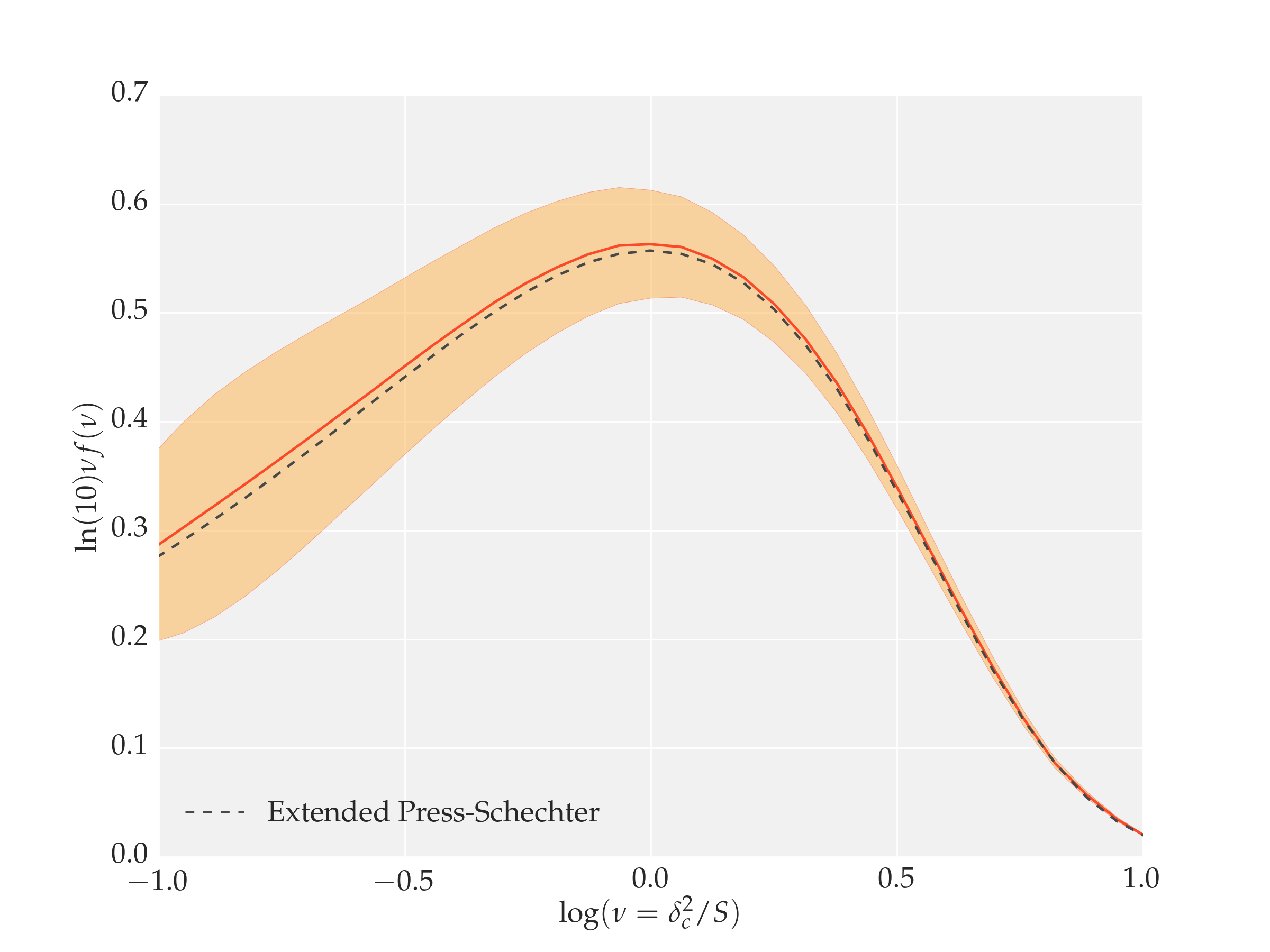}
\caption{The first up-crossing distribution re-stated in terms of height parameter is plotted versus $\delta_c^2/S$. The error bars are 1-$\sigma$ confidence level of the probability obtained from an ensemble of transition matrices introduced in Fig.(\ref{fig:WienerTra}).}
\label{fig:fu-mu}
\end{figure}

In Fig. \ref{fig:fu-mu}, we plot the the first up-crossing of the Wiener process $\nu f(\nu)$ redefined in terms of height parameter $\nu = \delta^2_c / S$ versus $\nu$ in the logarithmic scale.
The red solid line is the mean value of the $\nu f(\nu)$ and the orange shaded region corresponds to $1\sigma$ statistical error induced from transition matrix reconstruction. The black long dashed line is the prediction of the extended Press-Schechter formalism for first up-crossing amplitude.
In the next subsection we will investigate the evolution of the transition matrices and first up-crossing amplitude in dark sky simulation.

\subsection{Dark sky simulation and matrix formalism of EST}

In this subsection, we use the procedure explained beforehand to obtain the transition matrix using the Dark-Sky simulation. dark sky simulation is an N-body cosmological dark matter particle simulation to investigate the evolution of the structures in the Universe \cite{Skillman:2014qca}.

\begin{figure}[b]
\center
\includegraphics[width=0.5\textwidth]{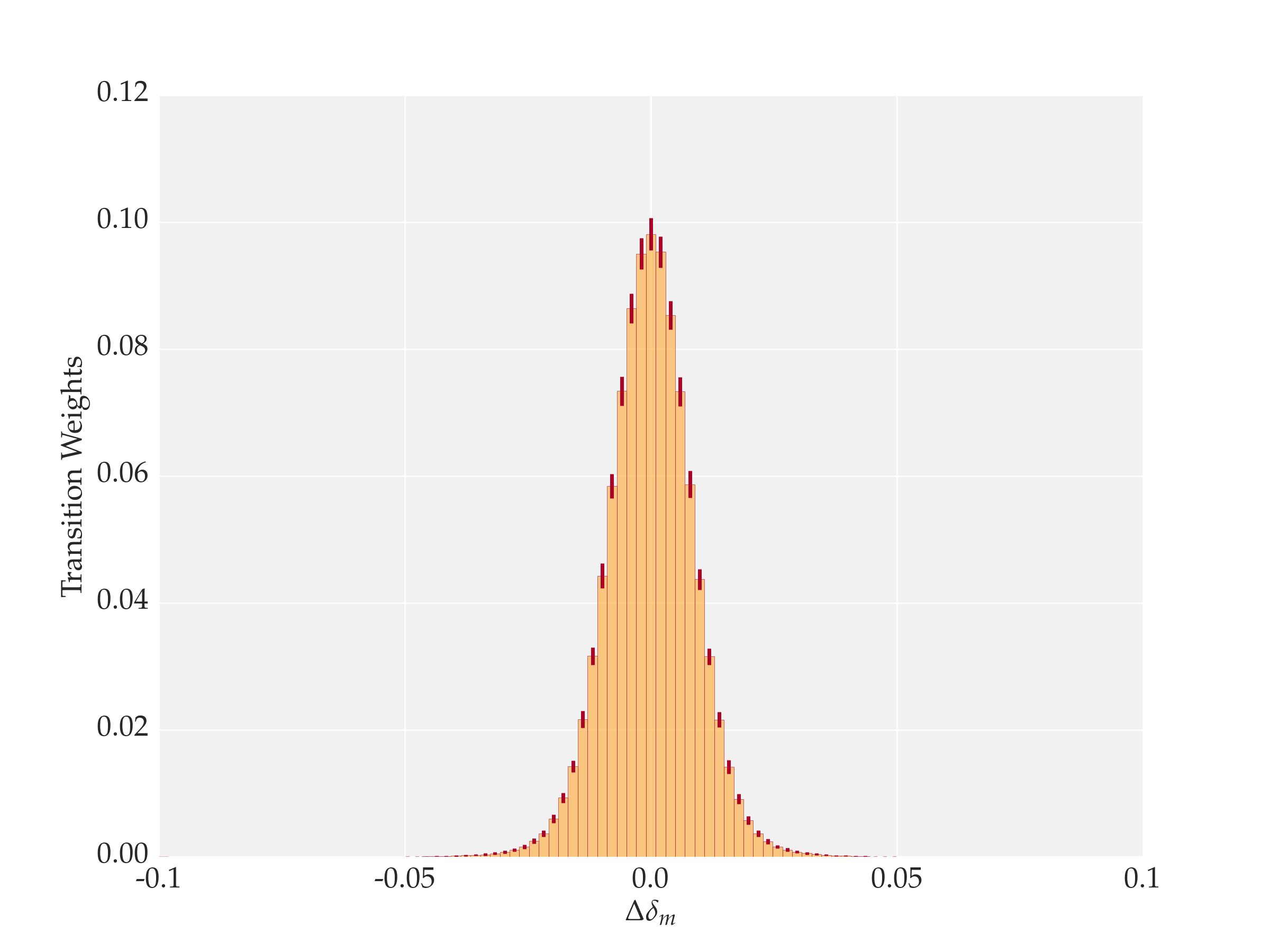}
\caption{{The normalized histogram of the transition in $\Delta\delta_m$ from dark sky simulation with a top-hat filter. The error bars were obtained from 1$\sigma$ confidence level from an ensemble of 100 series of $10^3$ trajectories.}}
\label{fig:THTra}
\end{figure}


\begin{figure}[h!]
\minipage{0.4\textwidth}
  \includegraphics[width=\linewidth]{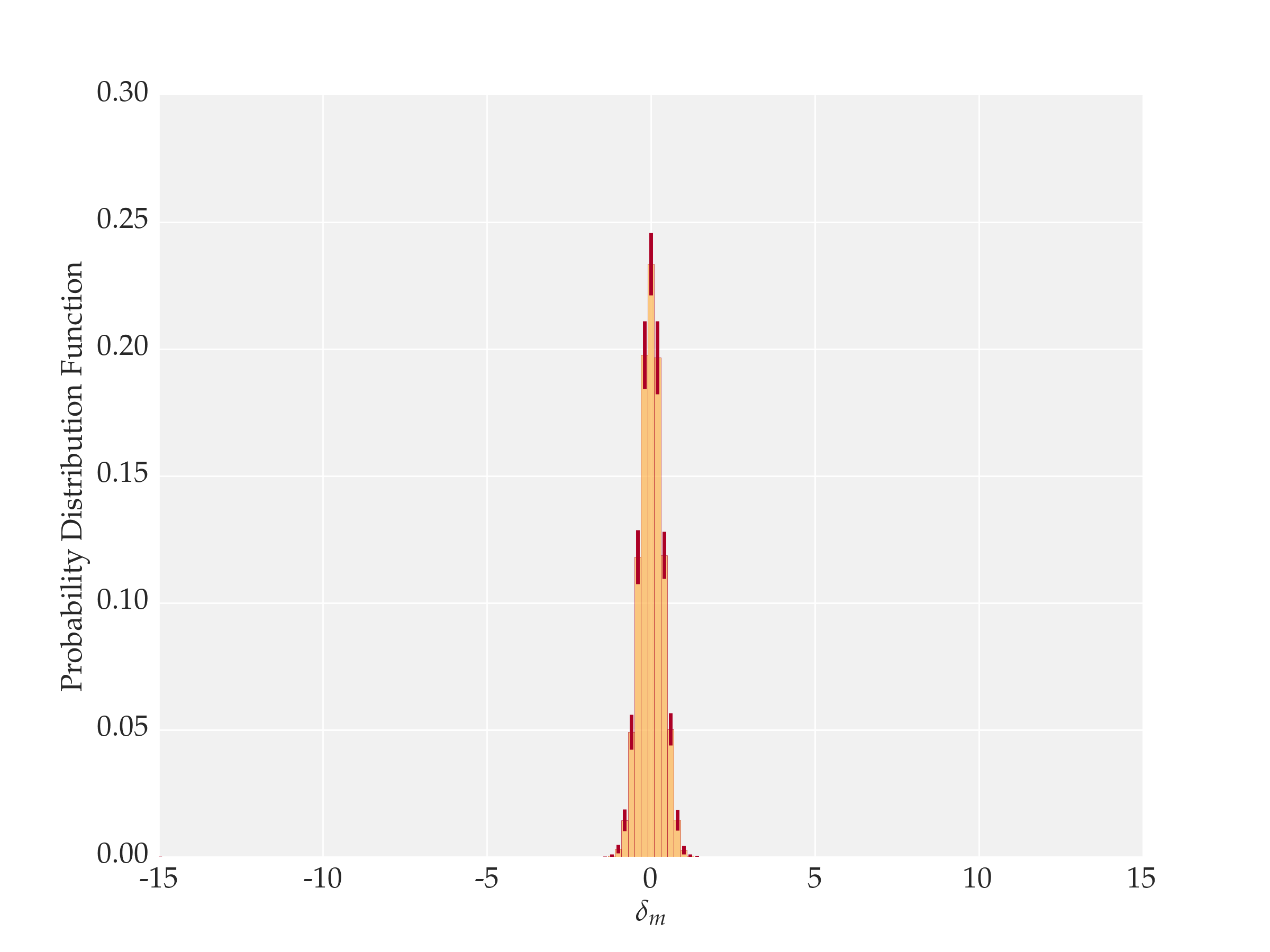}
\endminipage\hfill
\minipage{0.4\textwidth}
  \includegraphics[width=\linewidth]{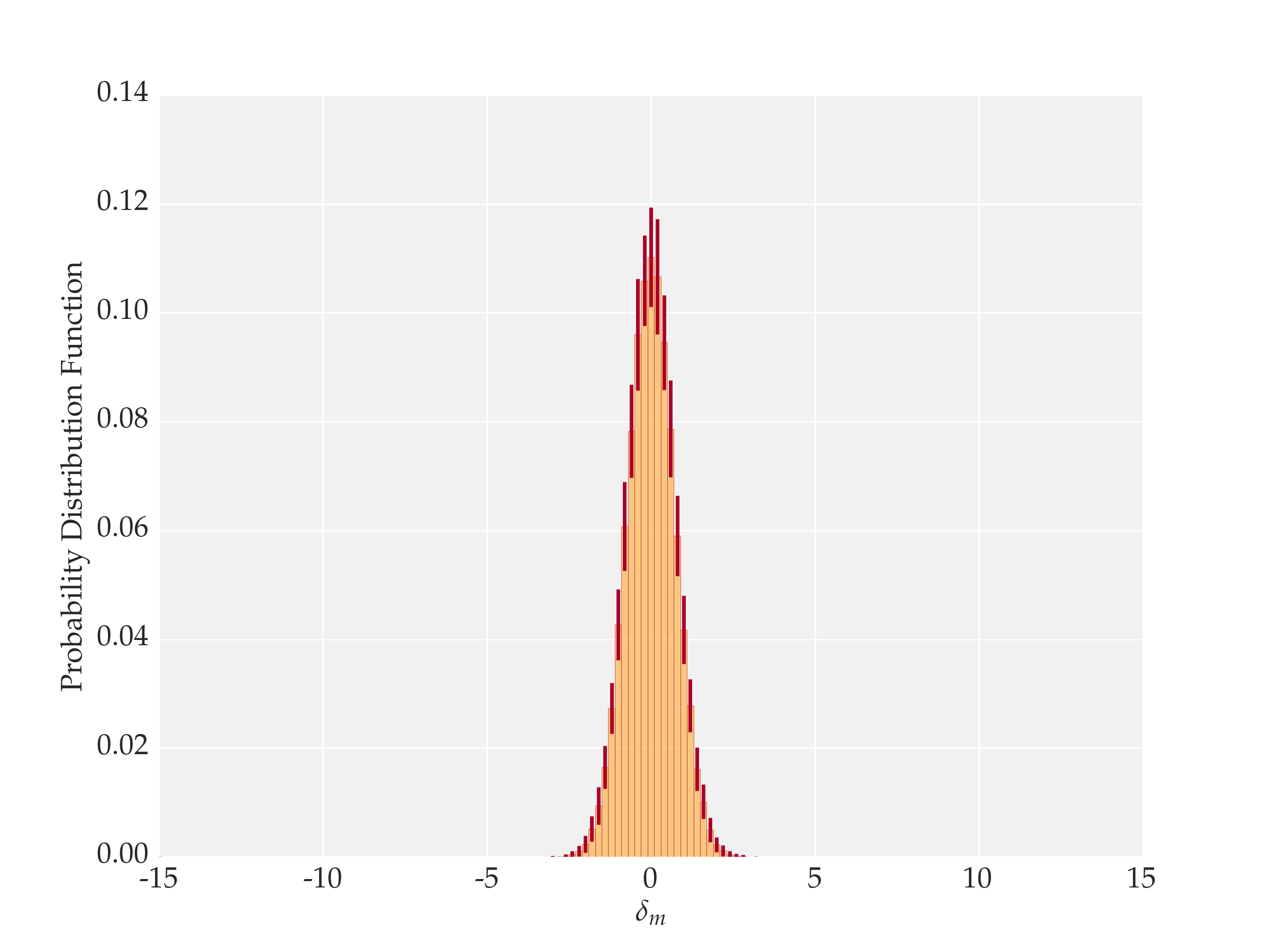}
\endminipage\hfill
\minipage{0.4\textwidth}%
  \includegraphics[width=\linewidth]{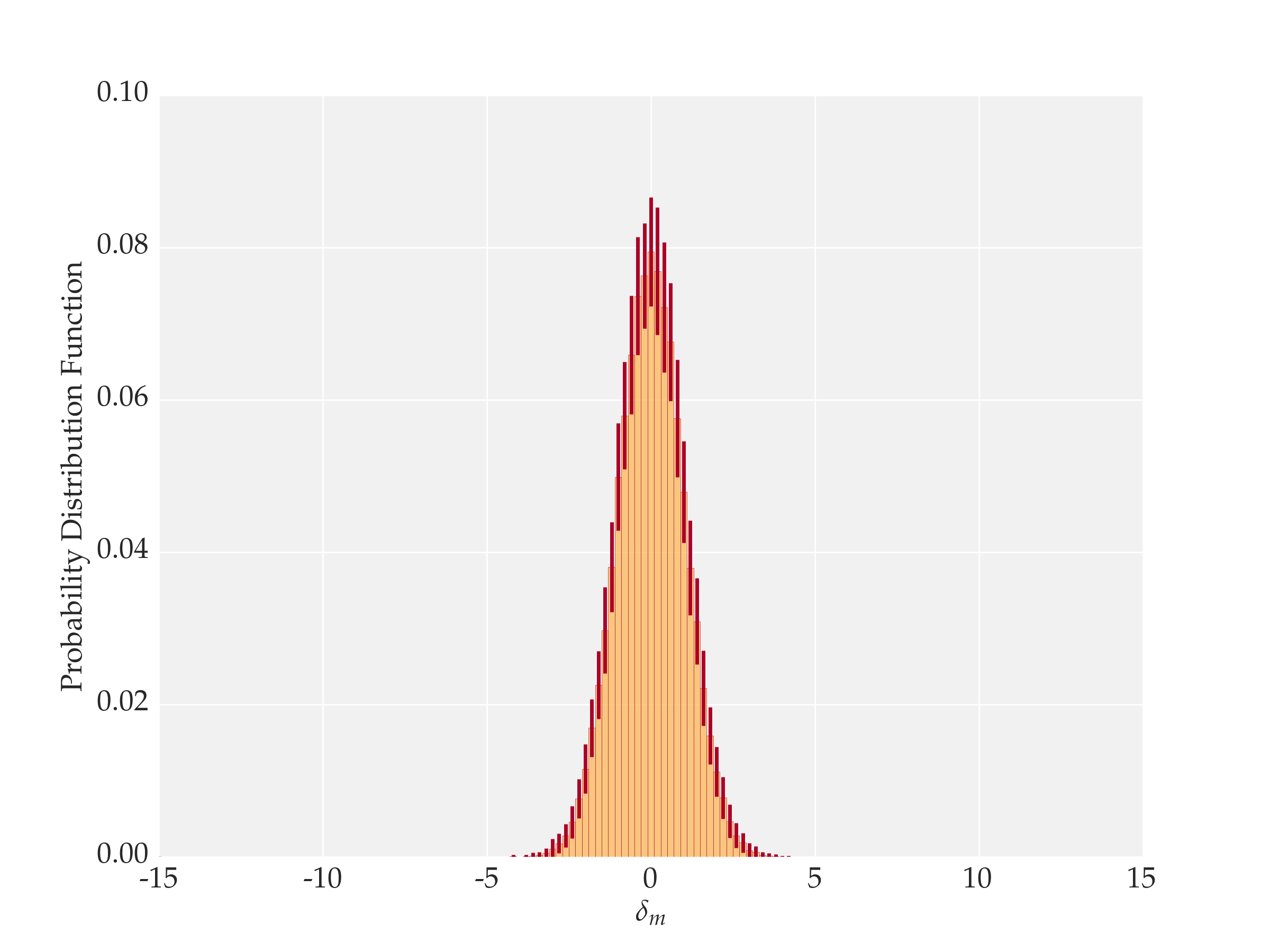}
\endminipage
\caption{From up to down, these figures show the normalized probability distribution function (PDF) of density contrast in three sequences of variances of $S=0.1$, $S=0.5$ and $S=1.0$ obtained from the dark sky simulation with a top-hat filter. The error bars are 1$\sigma$ confidence level of the probability obtained from an ensemble of transition matrices introduced in Fig. \ref{fig:THTra}.} \label{fig:THPS}
\end{figure}

In this study, we perform our analysis using The dark sky simulations data. The simulation is performed with different sizes and resolutions, all of which utilize the exact same $\Lambda$CDM cosmology with $(\Omega_m, \Omega_b, \Omega_\Lambda, h, \sigma_8)$ =
$(0.295, 0.0468, 0.705, 0.688, 0.835)$, and have box lengths from $100h^{-1}$Mpc to $8h^{-1}$Gpc respectively with $2048^3$ particles. To make the trajectories of EST, we used a 90Mpc box size of the simulation, then we used a snapshot of this simulation at $z=49$ and made 10,000 trajectories from different positions of comoving cosmological volume.

\begin{figure}[h]
\center
\includegraphics[width=\linewidth]{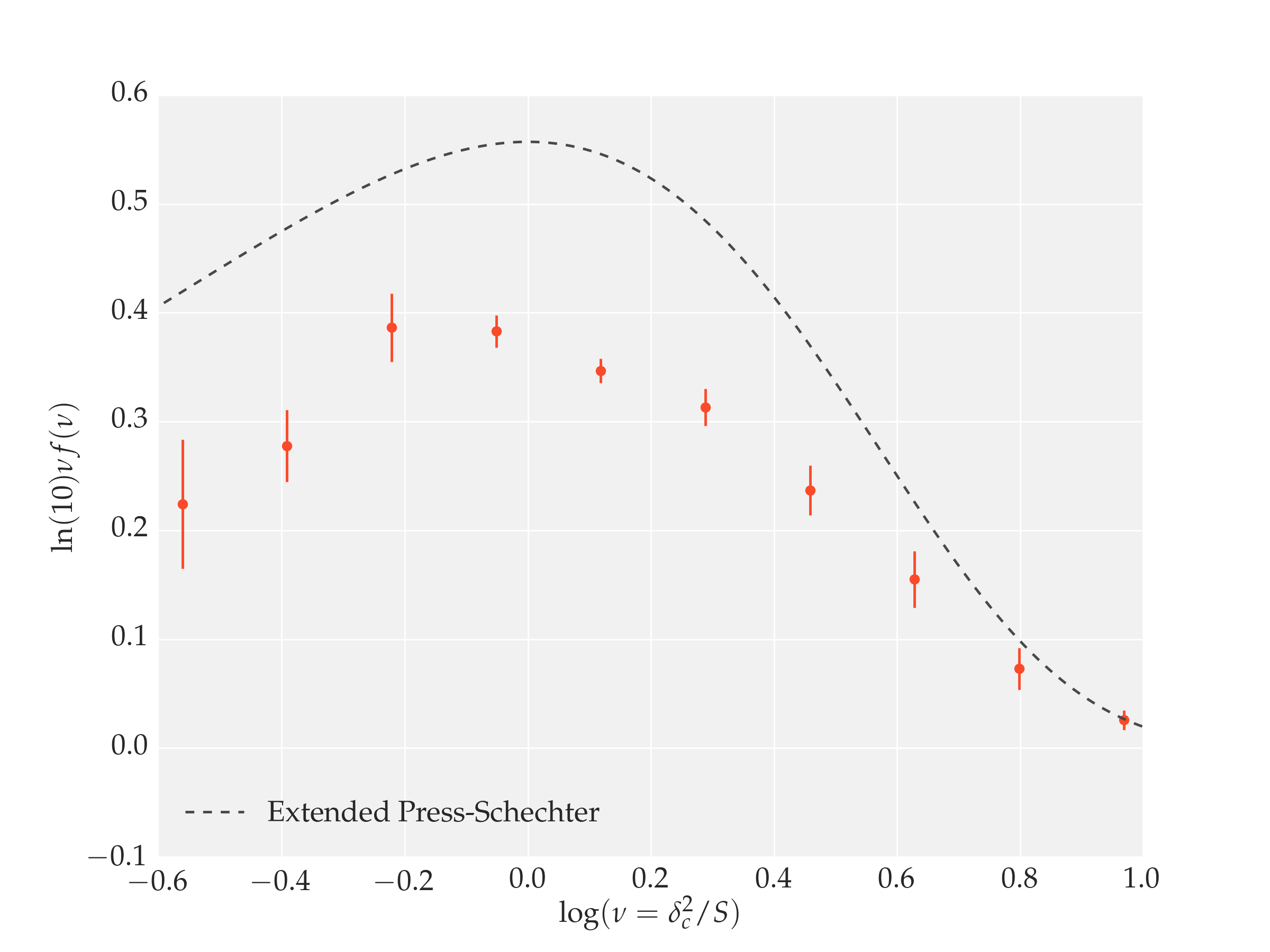}
\caption{The first up-crossing distribution restated in terms of height parameter is plotted versus $\delta_c^2/S$ for a top-hat filter used on dark sky simulation. The error bars are 1$\sigma$ confidence level of the probability obtained from an ensemble of transition matrices introduced in Fig.\ref{fig:THTra}.}
\label{fig:fu-muDS}
\end{figure}


In Fig.\ref{fig:THTra}, we plot the histogram of the transition matrix by the idea that the transition matrix is obtained by the histogram of $\Delta \delta_m$.
In Fig. \ref{fig:THPS} we plot the PDF of the density contrast in a sequence of variances. From up to down the variance is increased, which corresponds to smaller smoothing window functions. A very important point to note is that the trajectories are obtained by using the top-hat window function in real space. This means that the walks are correlated, where the standard solution of extended Press-Schechter will not work.
However, in the MF-EST we can find the first up-crossing distribution, by reconstruction of the transition matrix or just by using Eq.(\ref{Eq:f-fu}), where we have to construct the probability ket in each variance step. A very crucial point to indicate is that the probability ket is  equal to $n= S/\Delta S$ times the product of the transition matrix on the initial condition, however the transition matrix must be constructed by the histogram of $\Delta \delta_m$ where the steps of discretization $\Delta S _{d} \sim 0.17$ are larger than $\Delta S = 0.01$ which we use, in the definition of $n$. The $\Delta S _{d}$ as the new 
discretization length which is just defined for transition matrix construction is very similar to the Markov length (a length that makes the process looks like a Wiener one). In our case, this length is obtained by demanding that the probability ket from transition matrix construction must be the same as it is obtained from trajectories.
In Fig. \ref{fig:fu-muDS}, we plot the first up-crossing distribution in terms of height parameter $\nu f(\nu)$ versus $log (\delta_c^2/S)$ for the dark sky simulation where the error bars are extracted from the ensemble of trajectories we make from simulation using the top-hat filter. Our result is very similar to the results obtained by Musso and Sheth \cite{Musso:2013pha} for power law power spectrum by the method of back substitution.



\section{First up-crossing distribution of symmetric non-Gaussian field in Matrix formalism}
\label{sec:ng}

\begin{figure}[b]
\center
\includegraphics[width=0.5\textwidth]{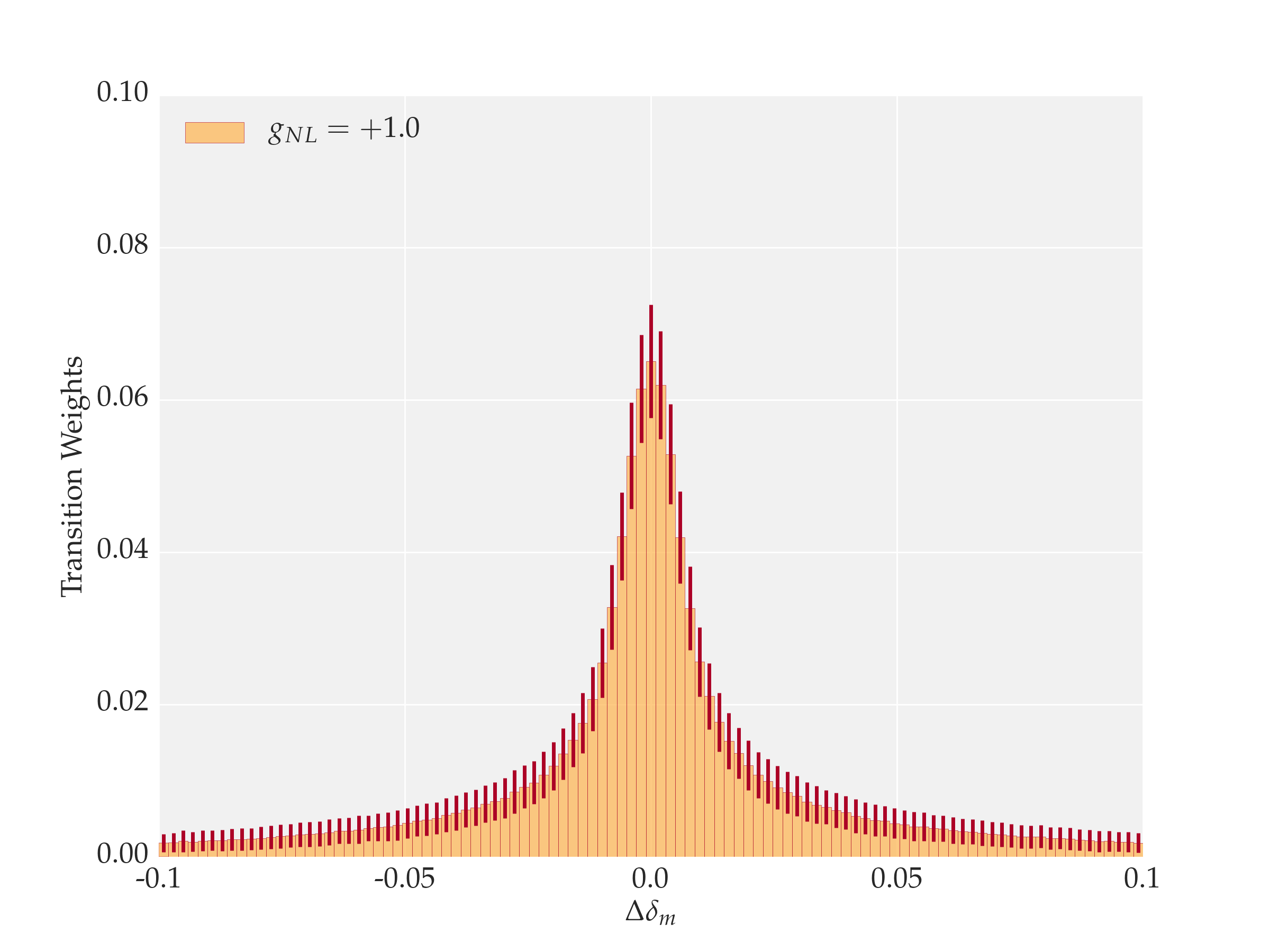}
\caption{The normalized histogram of the transition in $\Delta\delta_m$ for synthesized non-Gaussian density contrast with $g_{NL}$ type of non-Gaussianity. The error bars were obtained from  a 1$\sigma$ confidence level from an ensemble of 100 series of $10^3$ trajectories.}
\label{fig:TNG}
\end{figure}


\begin{figure}[h!]
\minipage{0.4\textwidth}
  \includegraphics[width=\linewidth]{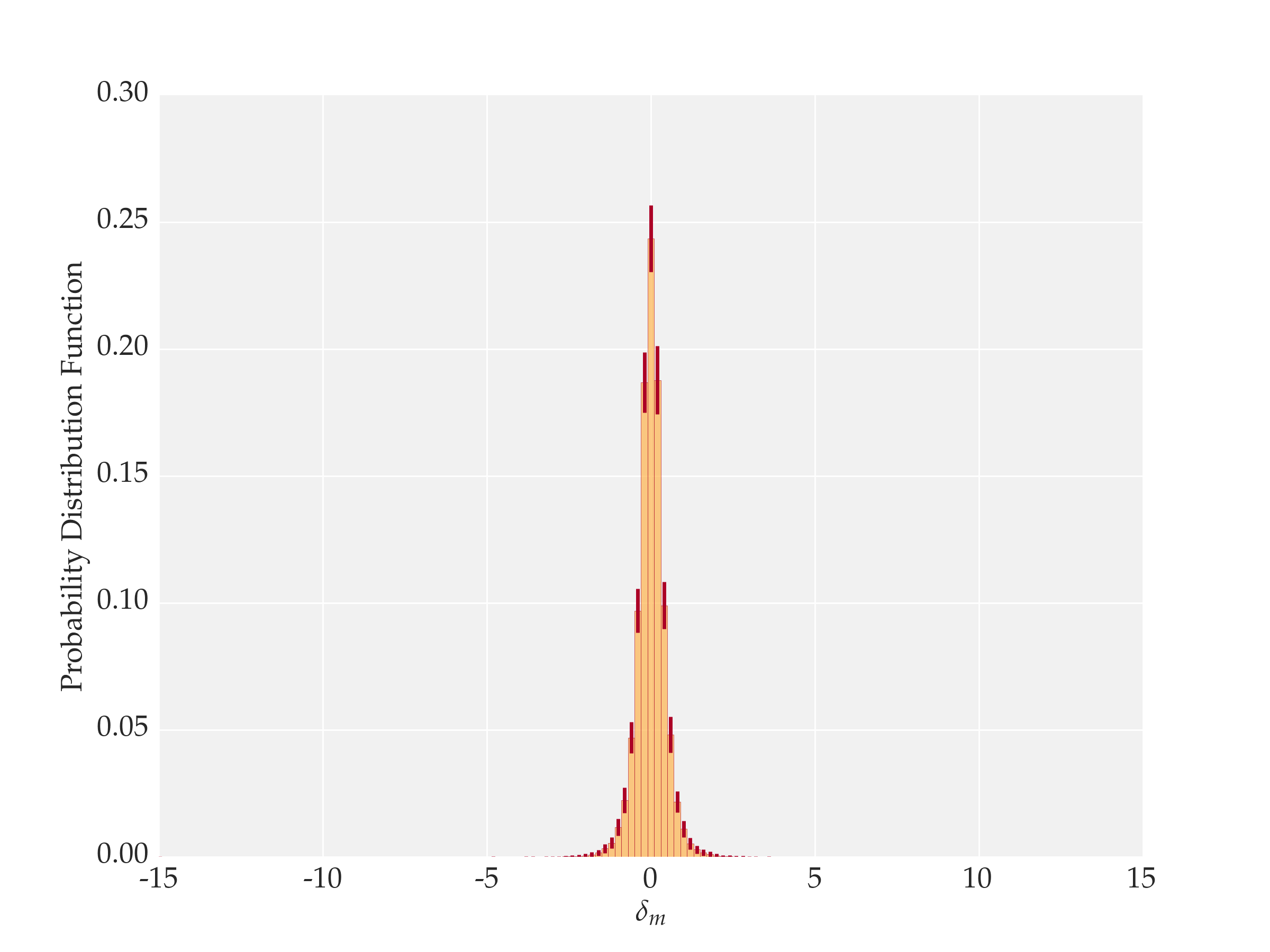}
\endminipage\hfill
\minipage{0.4\textwidth}
  \includegraphics[width=\linewidth]{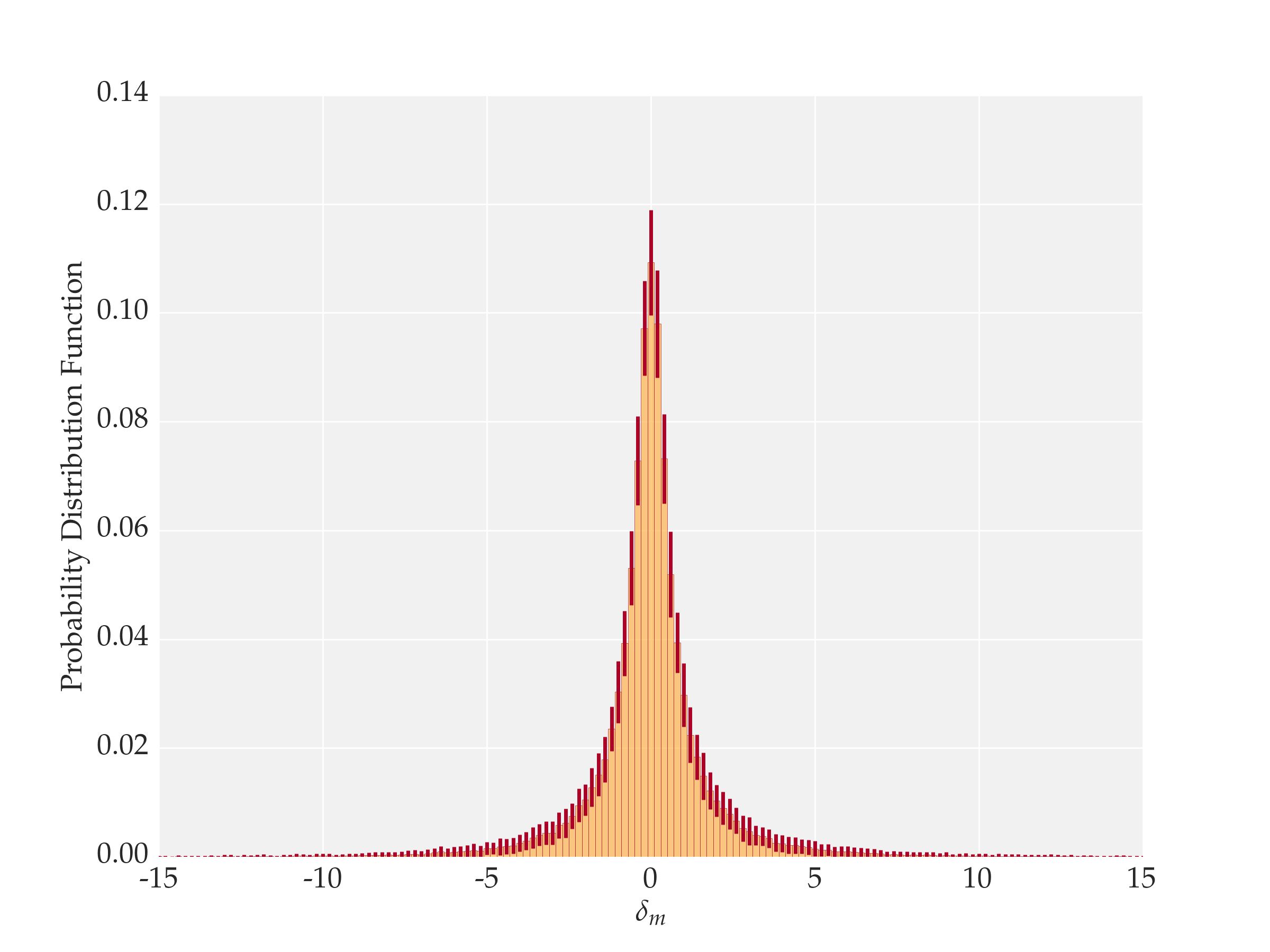}
\endminipage\hfill
\minipage{0.4\textwidth}%
  \includegraphics[width=\linewidth]{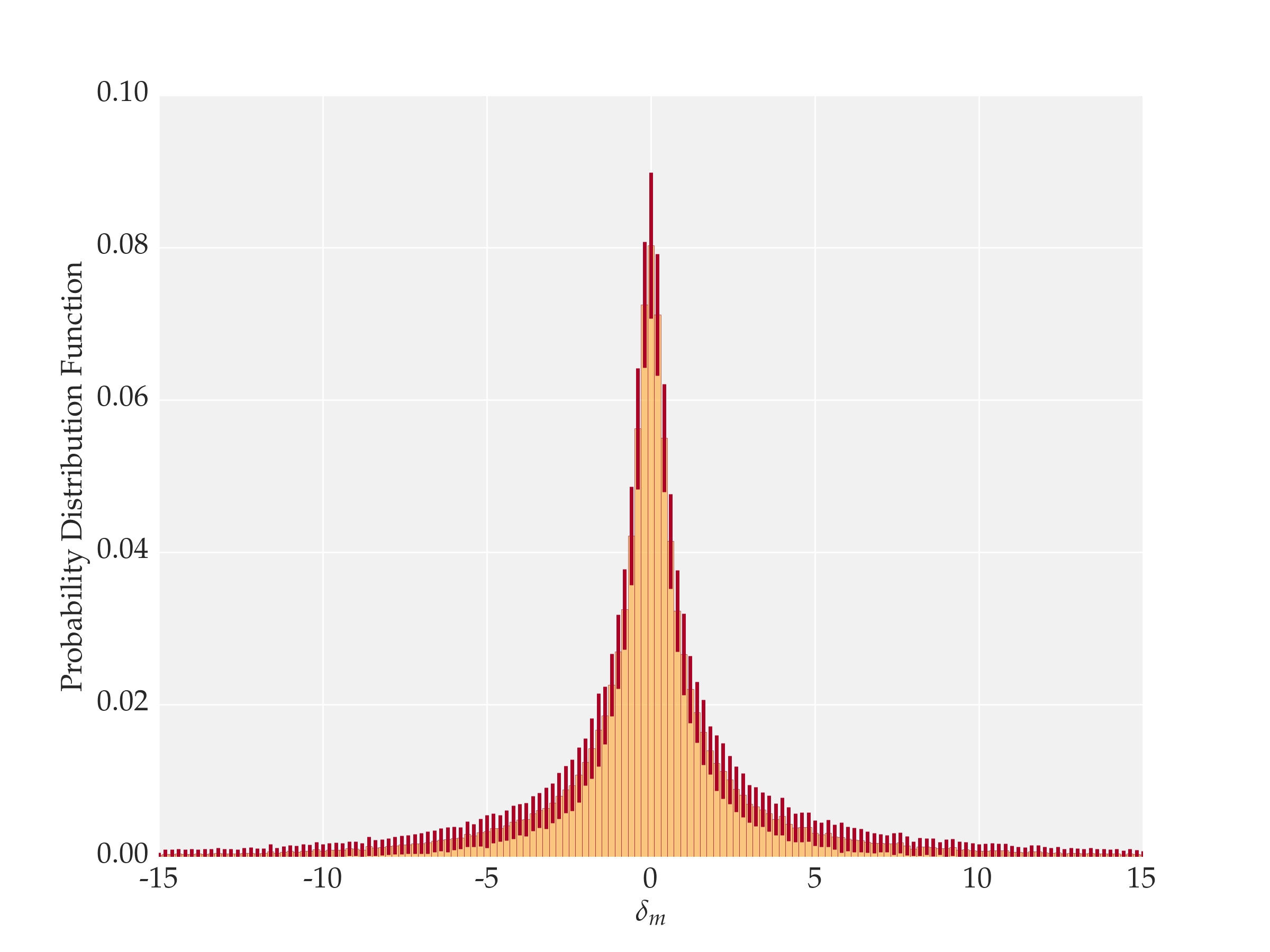}
\endminipage
\caption{From up to down, these figures show the normalized probability distribution function (PDF) of density contrast in three sequences of variances of $S=0.1$, $S=0.5$ and $S=1.0$ which were obtained from non-Gaussian distribution of initial field with $g_{NL}$ =1.0. The error bars were obtained from 1$\sigma$ confidence level of an ensemble of transition matrices introduced in Fig. \ref{fig:TNG}.} \label{fig:ps-ng}
\end{figure}

In this section, we extend the idea of matrix formalism for extraction of the first up-crossing distribution from a given trajectory ensemble, with a non-Gaussian distribution. However, we restrict ourselves to the symmetric non-Gaussian case. This is because the concept of mirror symmetry works in this type of models. In other words the processes that can be mapped to walks with zero drift can use the concept of mirror trajectory to solve the cloud-in-cloud problem.
As a specific example we study the $g_{NL}$ - type non-Gaussianity, which introduces a kurtosis. This type of non-Gaussianity is written in the form below
\be
\delta_m = \delta_g + g_{NL}\delta^3_{g}, \label{eq:gnl-ng}
\ee
where $\delta_g$ is the Gaussian density contrast and $g_{NL}$ is a constant.
We synthesize an ensemble of trajectories which  has a $g_{NL}$ type non-Gaussianity from a Wiener process, whose density contrast is mapped to a non-Gaussian one via Eq.(\ref{eq:gnl-ng}). Then we use the method of transition matrix construction by plotting the histogram of $\Delta \delta_m$.
In Fig. \ref{fig:TNG}, the normalized histogram of the transition in $\Delta\delta_m$ for synthesized non-Gaussian density contrast with $g_{NL}= +1$  as a specific example is plotted. The error bars were obtained from the 1$\sigma$ confidence level from an ensemble of 100 series of $10^3$ trajectories.
 Accordingly by the knowledge of transition matrix, the probability ket can be constructed for the non-Gaussian case. In Fig. \ref{fig:ps-ng}, we plot the probability ket, which is constructed from the transition matrix and initial probability ket.

\begin{figure}[h]
\center
\includegraphics[width=\linewidth]{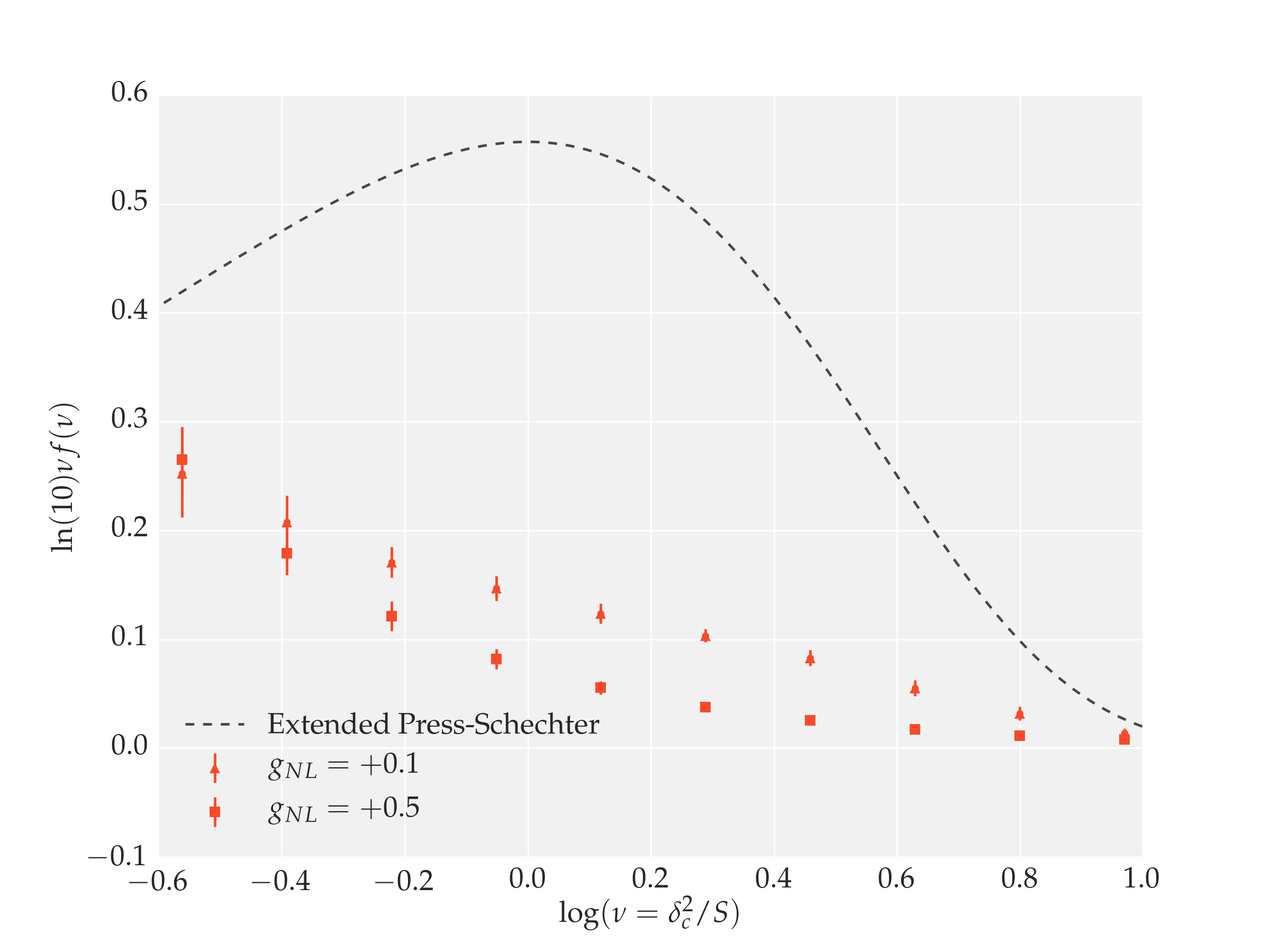}
\caption{The first up-crossing distribution restated in terms of height parameter is plotted versus $\delta_c^2/S$ for a non-Gaussian distribution of initial conditions (the triangles are for $g_{NL} = +0.1$ and the squares are for $g_{NL} =+ 0.5$). The error bars are 1$\sigma$ confidence level of the probability obtained from an ensemble of transition matrices introduced in Fig.\ref{fig:TNG}.}
\label{fig:fu-muNG}
\end{figure}

Finally in Fig.\ref{fig:fu-muNG}, the first up-crossing distribution restated in terms of height parameter is plotted versus $\delta_c^2/S$ for a non-Gaussian distribution of initial conditions with $g_{NL} = 0.1$ and $g_{NL} = 0.5$. The error bars are 1$\sigma$ confidence level of the probability obtained from an ensemble of transition matrices introduced in the histogram of $\Delta \delta_m$.
This section shows how the MF-EST can address the first up-crossing question with initial non-Gaussian conditions.
The extension of MF-EST for nonzero drifts can be an interesting extension to this work.
In the next section, we study the continuous limit of MF-EST as a consistency check which also introduces new concepts to study the nonlinear structure formation.

\section{Continuous limit  of Matrix formalism of EST}
\label{sec:cont}
In this section we consider the procedure to bring the matrix formalism of EST into a continuous limit manifestation. In the previous section, it was shown that the physics of the first up-crossing (number count of structures) is related to the variance derivative of the probability states and it was constructed by the probability transition rate matrix in discrete the 2D-EST plane. In the first subsection we derive the probability transition rate matrix in the continuous limit of matrix formalism in 2D-EST plane where the steps (variances) become continuous and the states (density contrasts) remain discrete. In the second subsection we derive the Fokker-Planck (FP) equation of dispersion and we show that this new formalism can reproduce all the structure of EST.
\subsection{Probability transition rate in continuous limit of matrix formalism EST}
In this subsection we want to find the probability transition rate in the continuous limit in variance. First of all we start with the definition of the variance derivative of probability ket
\begin{equation} \label{eq:cont}
	\frac{\partial{\Ket{\mathcal{P}_s}}}{\partial{S}}= \lim_{\Delta s \rightarrow 0} \frac{\Ket{\mathcal{P}_{s+\Delta s}} - \Ket{\mathcal{P}_s}}{\Delta S},\\
\end{equation}
where $\Delta S$ is the infinitesimal parameter in the variance axis.
Using the definition of the probability ket in Eq.(\ref{eq:ket}) and by adaption of more convenient notation ${\cal{P}}_s(\delta_m^i)\equiv{\cal{P}}(S,\delta_m)$ we will proceed.
The very crucial point is that the probability in step $S+ \Delta S$ can be written as a conditional one, in terms of the probability in the previous step (Markovian condition) as below:
\begin{equation} \label{eq:chap}
	\mathcal{P}(S + \Delta S, \delta_m ) = \displaystyle\sum_{\delta'_m} \mathcal{P}( S + \Delta S, \delta_m  | S,\delta'_m) \mathcal{P}(S,\delta'_m),
\end{equation}
\begin{widetext}
where the sum is over the density contrast value on the step $S$; substituting this in Eq.(\ref{eq:cont}) with the new notation, we will find
\begin{equation}
	\frac{\partial{\mathcal{P}(S,\delta_m)}}{\partial{S}} =\lim_{\Delta S\rightarrow 0} \frac{1}{\Delta S} \displaystyle\sum_{\delta'_m} \mathcal{P}(S,\delta'_m ) \left( \mathcal{P}(S+\Delta S, \delta_m| S, \delta'_m ) - \delta^k_{\delta_m, \delta'_m} \right),
\end{equation}
where $\delta^k_{\delta_m, \delta'_m}$ is the Kronecker delta. Since we want to take the limit $\Delta S \rightarrow 0$, we can expand the conditional probability $\mathcal{P}(S+\Delta S, \delta_m  | S, \delta'_m ) \equiv \mathcal{P}(\delta_m |\delta'_m ; \Delta S)$ in Taylor series and keep only the lowest term,
\begin{equation} \label{eq:cont1}
	\mathcal{P}(\delta_m |\delta'_m ; \Delta S) = \mathcal{P}(\delta_m |\delta'_m ; \Delta S)|_{\Delta S = 0} + \Delta S \frac{\partial \mathcal{P}(\delta_m |\delta'_m ; \Delta S)}{\partial \Delta S}|_{\Delta S = 0} + \dots
\end{equation}
We should note that the conditional probability is a function of $\Delta S$ and not the variance.
Now we analyze the two terms appearing in the rhs of Eq. (\ref{eq:cont1}).
The first term in the limit of  $\Delta S \rightarrow 0$ is as below:
\begin{equation}
	\mathcal{P}(\delta_m |\delta'_m ; \Delta S)|_{\Delta S = 0} = \delta^k_{\delta_m, \delta'_m}.
\end{equation}
For the second term appearing in the rhs of Eq. (\ref{eq:cont1}) we can define the continuous limit of probability transition rate $\mathcal{R}(\delta'_m , \delta_m ; \Delta S)$ as
\begin{equation}
\mathcal{R}(\delta'_m , \delta_m ; \Delta S) = \frac{\partial \mathcal{P}(\delta_m |\delta'_m ; \Delta S)}{\partial \Delta S}|_{\Delta S = 0}.
\end{equation}
We substitute these two terms into the Taylor series in Eq.(\ref{eq:cont1}); now Eq.(\ref{eq:cont}) can be written in terms of ${\cal{R}}(\delta'_m, \delta_m ;\Delta S)$ as
\begin{equation}
	\frac{\partial{\mathcal{P}( S , \delta_m)}}{\partial{S}} = \lim_{\Delta S \rightarrow 0} \frac{1}{\Delta S} \displaystyle\sum_{\delta'_m} \mathcal{P}(S, \delta'_m) \left( \delta^k_{\delta_m, \delta'_m} + \Delta S \mathcal{R}(\delta'_m, \delta_m ;\Delta S) - \delta^k_{\delta_m, \delta'_m} \right),
\end{equation}
\end{widetext}
\begin{equation}
	\frac{\partial{\mathcal{P}(S,\delta_m)}}{\partial{S}} = \displaystyle\sum_{\delta'_m} \mathcal{P}(S, \delta'_m) \mathcal{R}(\delta'_m, \delta_m ;\Delta S).
\end{equation}
The above equation is the continuous limit of matrix formalism of EST. Due to the definitions we introduced in the previous section in the matrix formalism of EST framework $\mathcal{P}(S,\delta_m) = \Braket{\delta_m|\mathcal{P}_s}$, $\mathcal{P}(\delta'_m, S) = \Braket{\delta'_m|\mathcal{P}_s}$ and $\mathcal{R}(\delta'_m, \delta_m ;\Delta S) = \Braket{\delta_m | \hat{\mathcal{R}} | \delta'_m}$, then Eq.(\ref{eq:cont}) will become
\begin{equation}
	\frac{\partial{\Braket{\delta_m|\mathcal{P}_s}}}{\partial{s}} = \displaystyle\sum_{\delta'_m} \Braket{\delta_m | \hat{\mathcal{R}} | \delta'_m} \Braket{\delta'_m|\mathcal{P}_s}= \Braket{\delta_m | \hat{\mathcal{R}} | \mathcal{P}_s},
\end{equation}
and finally this can be written in a more familiar way:
\begin{equation}
\boxed{\frac{\partial{\Ket{\mathcal{P}_s}}}{\partial{s}} = \hat{\mathcal{R}} \Ket{\mathcal{P}_s}}
\end{equation}
The continuous limit of the matrix formalism shows how consistent this formalism is with the standard case.
In the next subsection we will derive the Fokker-Planck equation in the continuous limit.

\subsection{Fokker-Planck equation in matrix formalism of EST}
In this subsection, we show that in the continuous limit of matrix formalism of EST we can find the Fokker-Planck (FP) equation.
First we define the density contrast difference $\Delta\delta_m = \delta_m - \delta'_m$  and rewrite Eq. ( \ref{eq:chap}),
\begin{equation}
\mathcal{P}( S + \Delta S,\delta_m) = \displaystyle\sum_{\Delta\delta_m} \mathcal{P}(\delta_m|\delta_m - \Delta\delta_m; \Delta s) \mathcal{P}(s,\delta_m - \Delta\delta_m),
\end{equation}
where we indicate that the conditional probability in the equation above does not depend on the variance. Now we can expand $\mathcal{P}(S,\delta_m - \Delta\delta_m )$ in terms of small $\Delta\delta_m$ up to second order as
\begin{widetext}
\begin{equation} \label{eq:deltataylor}
		\mathcal{P}(S + \Delta S, \delta_m ) = \displaystyle\sum_{\Delta\delta_m} \mathcal{P}(\delta_m|\delta_m - \Delta\delta_m; \Delta S) \bigg\{ \mathcal{P}(S,\delta_m) - \Delta\delta_m \frac{{\partial \mathcal{P}(S,\delta_m )}}{{\partial \delta_m}} + \frac{\Delta\delta_m^2}{2} \frac{\partial^2 \mathcal{P}(S,\delta_m)}{\partial \delta_m^2} \bigg\}. \nonumber
\end{equation}
By rearranging the terms in Eq.(\ref{eq:deltataylor}),
\begin{equation}
		\mathcal{P}(\delta_m, S + \Delta S) = \mathcal{P}(\delta_m, S)\displaystyle\sum_{\Delta\delta_m} \mathcal{P}(\delta_m|\delta'_m; \Delta S)
		- \frac{{\partial \mathcal{P}(\delta_m ,S)}}{{\partial \delta_m}} \displaystyle\sum_{\Delta\delta_m} \Delta\delta_m \mathcal{P}(\delta_m|\delta'_m; \Delta S) + \frac{1}{2} \frac{\partial^2 \mathcal{P}(\delta_m,S)}{\partial \delta_m^2} \displaystyle\sum_{\Delta\delta_m} \Delta\delta_m^2 \mathcal{P}(\delta_m|\delta'_m; \Delta S).
\end{equation}
\end{widetext}
It is interesting to note that second and third terms in the rhs represent the first and second moments of $\Delta\delta_m$, where by the knowledge that the mean and the variance is $\langle\Delta\delta_m\rangle = 0$ and $\langle\Delta\delta_m^2\rangle = \sigma_\Delta^2$, we will have
\begin{equation}
	\mathcal{P}(\delta_m, S + \Delta S) = \mathcal{P}(\delta_m, S) + \frac{1}{2} \frac{\partial^2 \mathcal{P}(\delta_m,S)}{\partial \delta_m^2} \sigma_\Delta^2.
\end{equation}
Accordingly the variance derivative of probability ket becomes
\small
\begin{equation}
	\lim_{\Delta S \rightarrow 0} \frac{\mathcal{P}(\delta_m, S + \Delta S) - \mathcal{P}(\delta_m, S)}{\Delta S} = \lim_{\Delta S \rightarrow 0} \frac{1}{2} \frac{\partial^2 \mathcal{P}(\delta_m,S)}{\partial \delta_m^2} \frac{\sigma_\Delta^2}{\Delta S}.
\end{equation}
\normalsize
The very interesting point  to indicate here that in the limit of $\Delta S\rightarrow 0$, the variance of transitions in density contrast becomes $\sigma^2_\Delta = \Delta S$. This is true, as mentioned before, when we have the sharp k-space filter (Markovian process).
Accordingly we rediscover the FP equation as below:
\begin{equation}
	\boxed{\frac{\partial \Ket{\mathcal{P}_s}}{\partial S} = \frac{1}{2}\frac{\partial^2 \Ket{\mathcal{P}_s}}{\partial \delta_m^2} }
\end{equation}
In this section, we use the continuous limit of the matrix formalism of excursion set theory to show two important ingredients of the formalism. First is the derivation of the probability transition rate matrix and the second is the FP equation. In the next section we conclude and will list the future prospects of this formalism.


\section{Most massive progenitors}
\label{sec:mmp}
One of the main questions in galaxy formation and evolution is the connection of the properties of the luminous matter with their dark matter host halos during its evolution. In this arena, the standard hierarchical structure formation indicates that the dark matter host halos are built up from small structures due to merger and mass accretion.

In this direction, the merger history of the dark matter halos plays a crucial role in building the history of dark matter assembly. EST in its own has a proposal to address the merger history of dark matter halos. For this question the main quantity that plays an important role is the probability of finding the most massive progenitors of a host dark matter halo. The most massive progenitor has the main mass contribution to building a dark matter halo. More precisely we can ask that if we observe a galaxy or a cluster of galaxies hosted by halo of mass $M_2$ in redshift $z_2$, what will be the probability of having a most massive progenitor of dark matter halo of mass $M_1$ in redshift $z_1$. To address this question we know from the formalism of EST that the redshift is implemented in the critical density barrier.
\begin{figure}[t!]
\center
\includegraphics[width=0.5\textwidth]{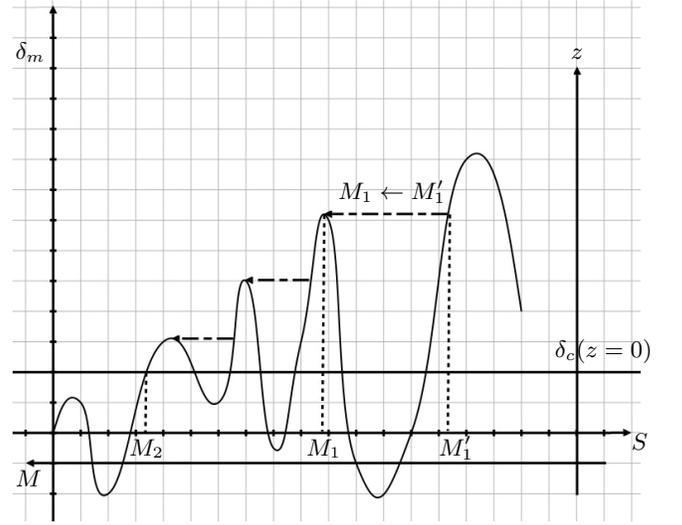}
\put(-245,175){$\delta_m$}
\put(-41,62){$\delta_c (z=0)$}
\put(-12,27){$S$}
\put(-245,13){$M$}
\put(-85,25){$M^\prime_1$}
\put(-135,25){$M_1$}
\put(-202,25){$M_2$}
\put(-123,122){$M_1 \leftarrow M^\prime_1$}
\put(-35,175){$z$}
\caption{A trajectory of a dark matter halo is plotted in the 2D-plane of EST. The horizontal dashed lines indicate the merging of the dark matter halo progenitors. The right vertical axes shows the redshift-dependent of the critical densities.}
\label{fig-mmp}
\end{figure}
In what follows we will reformulate the probability of massive progenitors in the language of MF of EST. Due to formalism we developed in previous sections, we can write $f(S_1, \delta_1 | S_2, \delta_2)$ as below:
\begin{equation}
f(S_1, \delta_1 | S_2, \delta_2) = - \frac{\partial \mathcal{F}_{(>s_1 | s_2)}}{\partial S_1} = - \frac{\partial}{\partial S_1} ( \Braket{\Delta | \hat{\mathcal{M}} | \mathcal{P}_{s_1 \vert s_2}}),
\end{equation}
where $\mathcal{F}_{(>s_1 | s_2)}$ is the fraction of trajectories that crossed $\delta_2$ at $S_2$ and their second barrier crossing is not before $S_1$. The probability ket $\Ket{{\mathcal{P}}_{s_1 \vert s_2}}$ is a conditional quantity that indicates the probability distribution of density contrast in $S_1$ where it has to pass the point $(S_2, \delta_2)$ in  2D plane of EST.
Now, we would like to calculate $\langle S_1 \rangle_{(\delta_1)}|_{(s_2, \delta_2)}$ which means the average of the variance $S_1$ with the conditional probability described above.
\begin{equation}
\begin{split}
	\langle S_1 \rangle_{(\delta_1)}&|_{(s_2, \delta_2)} = \int_{s_2}^{\infty} (S_1 - S_2)  f(S_1, \delta_1 | S_2, \delta_2) dS_1\\
	&= - \left\{ (S_1 - S_2)\mathcal{F}_{(>s_1 | s_2)}\Big|_{s_2}^{\infty}  - \int_{s_2}^{\infty} \mathcal{F}_{(>s_1 | s_2)} dS_1 \right\} \\
	&= \int_{s_2}^{\infty} \Braket{\Delta | \hat{\mathcal{M}} | \mathcal{P}_{s_1\vert s_2}} dS_1.
\end{split}
\end{equation}
So we will have
\begin{equation}
\boxed{
\begin{split}
	\langle S_1 \rangle_{(\delta_1)}|_{(s_2, \delta_2)} &= \int_{s_2}^{\infty}\int_{-\infty}^{\delta_1} \mathcal{P}(S_1, \delta_1 | S_2, \delta_2) d\delta_1 dS_1 \\
	&- \int_{s_2}^{\infty}\int_{\delta_1}^{\infty} \mathcal{P}(S_1, \delta_1 | S_2, \delta_2) d\delta_1 dS_1
\label{Eq:s-avg}
\end{split}}
\end{equation}
where mean and variance of the conditional probability are
\begin{equation}
\begin{split}
	\mu_{s_1, \delta_1 | s_2, \delta_2} &= \langle \delta_1 \delta_2 \rangle \frac{\delta_2}{S_2}, \\
	\sigma^2_{s_1, \delta_1 | s_2, \delta_2} &= S_1 - \frac{\langle \delta_1 \delta_2 \rangle^2}{S_2}.
\end{split}
\end{equation}
The quantity which is calculated in Eq.(\ref{Eq:s-avg}) can be interpreted as the average of most massive progenitor mass in specific redshift $z_1 \equiv \delta_1$; this can be illuminated by Fig. \ref{fig-mmp}. This figure shows the trajectory of a dark matter halo with mass $M_2$ in the present time. The crossing of the redshift-dependent critical density barrier with the trajectory shows the specific redshift of interest (right vertical axes) and the corresponding substructure masses. We should also note, that the x-axes is identical to the mass of dark matter halos (in inverse direction to the variance axes) and their substructures. $\langle S_1 \rangle_{(z_1)}|_{(s_2, \delta_c)}$ is the average over all realizations of merger histories with the condition that a subhalo of mass $M_1$ in redshift $z_1$ is a progenitor of a dark matter halo with mass $M_2$ in the present time. This section shows how naturally MF-EST is capable of finding the average function of most massive progenitor masses versus redshift. In the next section, we conclude and we will discuss some prospects of MF-EST.

\section{Future Prospects and Conclusion}
\label{sec:conclusion}
The excursion set theory is a formalism to study nonlinear structure formation. Its main goal is to answer the questions about the distribution of first up-crossings and eventually the distribution of structures in the Universe.
In this work we introduce a matrix formalism of EST. The first step is the discretization of the 2D plane of EST, where the trajectories become a discrete jumps  in the variance-matter density contrast plane. In this formalism we encapsulate the characteristics of the EST theory like the Gaussianity of the steps and the properties of the mirroring trajectories in 2D plane of EST via the transition matrix. \\
In the simplest case of the  Markovianity condition we show that all we need to know to construct the probability ket in a specific size window function (sharp k-space filter) is the initial probability ket and the transition matrix. We showed that a series of $n = S / \Delta S$  Gaussian transitions, will lead to a Gaussian profile for density contrast with a variance $S$. This result is identical to the PS formalism assumption for density contrast. We also discuss the statistics in EST, where we propose a procedure to find the cross correlation of any desired  quantities in 2D plane of EST.  We also discuss the number count in the discretized version of EST. For this we define the block matrix ${\hat{\cal{M}}}$. The matrix product of  ${\cal{\hat{M}}}$ with the probability ket selects the desired portion of trajectories which is necessary for barrier crossing and also excluding the cloud-in-cloud problem.
Accordingly by knowing the transition matrix and initial probability ket, we assert that by choosing the sharp k-space filter we can find the number density of structures. This is one of the practical results of the matrix formalism of EST, which can be applied to real/simulated data. To show this use the dark sky N-body simulation to extract the first up-crossing distribution which is constructed by the transition matrix. We argue that in N-body simulation we use the top-hat filter to construct the walks. In this case the correlation of steps and non-Markovianity emerges. Then we show that the discretization helps us to make the process look like a Wiener one, where the probability kets can be obtained by the transition matrix or directly from trajectories. The distribution of the first up-crossing with top-hat filter is obtained accordingly. As another specific example we discuss the distribution of first up-crossing with $g_{NL}$-type non-Gaussianity which can be considered as a zero drift process. In this direction we discuss and show how MF-EST is capable to find the average of the most massive progenitor masses in a fixed redshift. Finally, in this work the continuous limit of the EST in the probability ket is discussed and we consistently show that the matrix formalism EST goes to its original version as the Fokker-Planck equation is recovered.\\
As a future prospect, this work can be extended to incorporate the concept of voids and two barrier crossing \cite{Sheth:2003py,Paranjape:2011bz}, moving barriers \cite{Musso:2012qk}, primordial non-Gaussianity with nonzero drift \cite{Afshordi:2008ru,LoVerde:2011iz} and the halo bias \cite{Sheth:1999mn}. As an another direction the transition matrix, probability transition rate matrix and the number count procedure can be applied to the dark matter N-body simulations to reconstruct the merger history.
Eventually we want to indicate that the matrix formalism of EST will open up a new horizon for nonlinear structure formation and introduce matrices in this type of studies which are computationally more desirable.

\acknowledgments
S.B. acknowledges the hospitality of the Abdus Salam International Centre for Theoretical
Physics (ICTP) during the final stage of this work.
We should thank  Kasra Alishahi, Saeed Mahdisoltani, Laleh Memarzadeh, and Abolhassan Vaezi for extensive discussions on the idea of matrix formalism of EST.
We would like to thank  Ali Akbar Abolhasani, Nima Khosravi, Mohammad Reza Rahimi Tabar, Sohrab Rahvar, and Ravi Sheth for insightful comments.


\appendix
\section{ Statistics in matrix formalism of EST}
\label{sec:specific}

This Appendix describes some of the properties of EST in matrix formalism and we want to show the capability of this formalism in calculating the statistical quantities. In this arena, we study the concept of the expectation value and cross correlation of observables.
First of all, in order to find the expectation value of any desired quantity as a function of the density contrast we use the adder Delta  $\displaystyle \sum_{j} \Bra{\delta_m^j} = \Bra{\Delta}$ as a useful tool for our calculations, keeping in mind the relation introduced in Eq.(\ref{eq:Delta})
and the theorems we showed in the previous section, we have $\Bra{\Delta}\hat{\mathcal{T}} = \Bra{\Delta}$. This relation will help us to find the expectation value of $\mathcal{F}_s (\delta_m)$ as a function of $\delta_m$ as below:
\begin{equation}
	\hat{\mathcal{F}_s} =
	\left(\begin{smallmatrix}
		{\mathcal{F}_s (\delta_m^1)}\\
		&{\mathcal{F}_s (\delta_m^2)}\\
		&&\ddots\\
		&&&{\mathcal{F}_s (\delta_m^n)}
	\end{smallmatrix}\right).
\end{equation}
It is obvious that we can write $\Braket{\delta_m^i| \hat{\mathcal{F}}_s | \delta_m^j}$ in terms of the matrix components defined above:
\begin{equation}
	\Braket{\delta_m^i| \hat{\mathcal{F}}_s | \delta_m^j} = \mathcal{F}_s (\delta_m^i) \delta_{ij}. \\
\end{equation}
By knowing $\hat{\mathcal{F}}_s \Ket{\delta_m^i} = \mathcal{F}_s (\delta_m^i) \Ket{\delta_m^i}$ and $\Bra{\delta_m^i}\hat{\mathcal{F}}_s = \mathcal{F}_s (\delta_m^i) \Bra{\delta_m^i}$ we can write down the expectation value as
\begin{eqnarray}
	\Braket{\mathcal{F}_s (\delta_m)} &=& \displaystyle\sum_{i} \mathcal{F}_s (\delta_m^i) \mathcal{P}_s (\delta_m^i) \nonumber \\
&=&\Braket{\Delta | \hat{\mathcal{T}}^{-s} \hat{\mathcal{F}}_s \hat{\mathcal{T}}^s | \mathcal{P}_0}.\label{eq:exp1}
\end{eqnarray}
The first line of Eq.(\ref{eq:exp1}) is by definition of the probability ket. The second equality comes up from the definition of the expectation value of ${\cal{F}}$. Then we used the definition of the adder Delta and by the assumption of the Markovianity, we write the probability in the specific step $S$ in terms of transition matrix applied to probability ket. Finally, now we can define $\hat{{F}}_s$ as below to write the expectation value in a sophisticated way. This is possible because of the relation between the transition matrix and the adder Delta $\Bra{\Delta}$,
\begin{equation}
\Braket{\mathcal{F}_s (\delta_m)} = \Braket{\Delta | \hat{F}_s | \mathcal{P}_0},
\end{equation}
where $\hat{{F}}_s = \hat{\mathcal{T}}^{-s} \hat{\mathcal{F}}_s \hat{\mathcal{T}}^s $. This is, schematically,  very similar to the case of the Heisenberg picture of operators in quantum mechanics.
Now that we found a general expression for the expectation value, the next important step in statistics is the cross correlation functions of density contrast tracers. Below we will compute the correlation of two different functions like $\mathcal{F}_s$ and $\mathcal{G}_{s'}$ at two different sizes $S , S'$, where we set $S > S'$:
\small
\begin{equation}
	\Braket{\mathcal{F}_s (\delta_m) \mathcal{G}_{s'} (\delta_m)} 
	= \displaystyle\sum_{i,j} \mathcal{F}_s (\delta_m^i) \mathcal{G}_{s'} (\delta_m^j) \mathcal{P} (\delta_m^i , s | \delta_m^j , s') \mathcal{P} (\delta_m^j , s'),
\end{equation}
\normalsize
where we used the conditional probability to express the joint term $ \mathcal{P} (S,\delta_m^i  ; S' , \delta_m^j )$. The conditional probability can be expressed in terms of transition matrix corresponding to the steps $S$ and $S'$,
\begin{equation}
	\mathcal{P} (S,\delta_m^i | S', \delta_m^j ) = \braket{\delta_m^i| \hat{\mathcal{T}}^{s-s'} | \delta_m^j },
\end{equation}
then the cross correlation of two functions ${\mathcal{F}}_s (\delta_m)$ and ${\mathcal{G}}_{s'} (\delta_m)$ will become
\small
\begin{eqnarray}
\Braket{\mathcal{F}_s (\delta_m) \mathcal{G}_{s'} (\delta_m)} &=& \displaystyle\sum_{i,j} \mathcal{F}_s (\delta_m^i) \mathcal{G}_{s'} (\delta_m^j) \braket{\delta_m^i| \hat{\mathcal{T}}^{s-s'} | \delta_m^j} \Braket{\delta_m^j | \mathcal{P}_{s'}} \nonumber \\
	&=& \braket{\Delta | \hat{\mathcal{T}}^{-s} \hat{\mathcal{F}_s}\hat{\mathcal{T}}^s \hat{\mathcal{T}}^{-s'} \hat{\mathcal{G}_{s'}} \hat{\mathcal{T}}^{s'}| \mathcal{P}_0},
\end{eqnarray}
\normalsize
where we relate the probability ket to the transition matrix and initial probability ket.
Similar to the previous section we define $\hat{F}_s$ and $\hat{G}_{s'}$ as below
\begin{equation}
	\hat{F}_s= \hat{\mathcal{T}}^{-s} \hat{\mathcal{F}_s}\hat{\mathcal{T}}^s, \hspace{40pt} \hat{G}_{s'} = \hat{\mathcal{T}}^{-s'} \hat{\mathcal{G}_{s'}} \hat{\mathcal{T}}^{s'},
\end{equation}
which means that the cross correlation of two quantities in EST can be obtained by the initial probability ket, adder Delta $\Bra{\Delta}$ and the operators
$\hat{F}_s$ and $\hat{G}_{s'}$ as below
\begin{equation} \label{eq:cross}
\Braket{\mathcal{F}_s (\delta_m) \mathcal{G}_{s'} (\delta_m)} = \braket{\Delta | \hat{F}_s \hat{G}_{s'}| \mathcal{P}_0}.
\end{equation}
Equation(\ref{eq:cross}) is the important equation which can be used to calculate each correlation function practically in N-body simulations.
\end{document}